\pdfoutput=1
\documentclass[conference]{IEEEtran}
\IEEEoverridecommandlockouts

\usepackage{cite}
\usepackage{amsmath,amssymb,amsfonts}
\usepackage{amsthm}
\newtheorem{theorem}{Theorem}
\usepackage{algorithm}
\usepackage{algpseudocode}
\usepackage{graphicx}
\usepackage{textcomp}
\usepackage{xcolor}
\usepackage{xspace}
\usepackage{stmaryrd}
\usepackage{caption}
\usepackage{subcaption}
\usepackage{booktabs}
\usepackage{multirow}
\usepackage{tikz}
\usepackage{hyperref}
\usepackage{xurl}
\definecolor{linkpurple}{cmyk}{0.55,1,0,0.15}
\definecolor{linkblue}{cmyk}{1,0.58,0,0.21}
\hypersetup{breaklinks=true,colorlinks,linkcolor=linkpurple,citecolor=linkpurple,urlcolor=linkblue}

\DeclareRobustCommand\numcircledtikz[1]{\tikz[baseline=(char.base)]{
    \node[shape=circle,draw,fill,inner sep=1pt] (char)
    {\textcolor{white}{#1}};}}

\newenvironment{packeditemize}{
\begin{list}{$\bullet$}{
\setlength{\itemsep}{0pt}
\addtolength{\labelwidth}{10pt}
\setlength{\leftmargin}{12pt}
\setlength{\listparindent}{\parindent}
\setlength{\parsep}{2pt}
\setlength{\topsep}{0pt}}}
{\end{list}
}

\def\Snospace~{\S{}}

\newcommand{\ie}{{i.e.,}~}
\newcommand{\eg}{{e.g.,}~}

\newcommand{\sys}{{\textsf{Reforge}}\xspace}
\newcommand{\syssrpe}{{\textsf{Reforge~(SRPE)}}\xspace}
\newcommand{\syscgp}{{\textsf{Reforge~(CGP)}}\xspace}
\newcommand{\dglf}{{\textsf{DGL~(Full)}}\xspace}
\newcommand{\dglns}{{\textsf{DGL~(NS)}}\xspace}

\begin{document}

\title{\sys: Low-Latency Distributed GNN Serving with Selective Embedding Recomputation\\
{\large Extended version of the IPDPS 2026 paper~\cite{reforge_ipdps26}}}

\author{
\IEEEauthorblockN{Geon-Woo Kim, Donghyun Kim, Jeongyoon Moon, Henry Liu, Tarannum Khan,\\
Anand Iyer\textsuperscript{$\dagger$}, Daehyeok Kim, Aditya Akella}
\IEEEauthorblockA{\textit{The University of Texas at Austin}, \textit{\textsuperscript{$\dagger$}Georgia Institute of Technology}\\
\{gwkim, donghyun, jeongyoonm, henry.liu, tarannum.khan, daehyeok\}@utexas.edu,\\
anand.iyer@gatech.edu, akella@cs.utexas.edu}
}

\maketitle
\pagestyle{plain}
\thispagestyle{plain}

\begin{abstract}
Graph Neural Networks (GNNs) have been widely adopted for their ability to compute expressive node representations in graph datasets. However, serving GNNs on large graphs is challenging due to the high communication, computation, and memory overheads of constructing and executing computation graphs, which represent information flow across large neighborhoods. Existing approximation techniques in training can mitigate the overheads but, in serving, still lead to high latency and/or accuracy loss.
To this end, we propose \sys, a system that enables low-latency GNN serving for large graphs with minimal accuracy loss through two key ideas. First, \sys employs \emph{selective recomputation of precomputed embeddings}, which allows for reusing precomputed computation subgraphs while selectively recomputing a small fraction to minimize accuracy loss. Second, we develop \emph{computation graph parallelism}, which reduces communication overhead by parallelizing the creation and execution of computation graphs across machines.
Our evaluation with large graph datasets and GNN models shows that \sys significantly outperforms state-of-the-art techniques.
\end{abstract}

\begin{IEEEkeywords}
Graph Neural Networks, Low-Latency Serving, Efficient Inference, Parallelization Technique
\end{IEEEkeywords}

\section{Introduction}\label{sec:intro}
Graph Neural Networks (GNNs) have gained widespread
attention due to their ability to provide breakthrough results in diverse
domains~\cite{wu2020graph,fout2017protein,zhou2020graph,GNNSocialRec,GNNCommunicationNet,DeepTMA}; they have been instrumental in discovering life-saving
drugs~\cite{Lo2018,Stokes2020}, empowered recommendation systems~\cite{PinSage2018,MAG2015}, and helped in traffic planning~\cite{Deng2019,jiang2022graph,gnntraffic1}.
Recent studies have enabled reasoning on graph-structured data~\cite{GreaseLM2021,Graphformers2021,Gnp2024,ReLM2023,Prot2text2024} by enhancing large language models (LLMs)~\cite{achiam2023gpt,gpt3,wei2022emergent,hoffmann2022training} with GNNs to effectively capture intricate structural information.

The GNN lifecycle consists of two phases: \emph{training} and \emph{serving}. In the training phase~\cite{P3,DistDGL2020,PyG2019,NeuGraph2019,PaGraph2020,DGL,GraphNets}, GNNs take as input graph-structured data and feature vectors of each node (\autoref{fig:background1} (left)), and learn the \emph{embeddings} of each
node in the graph. In the serving phase~\cite{quiver2023, PinSage2018, lambdagrapher, fograph,lin2022platogl}, the trained model is used to predict the embeddings for previously unseen \emph{query} nodes.
Such predictions empower downstream GNN use-cases, such as malicious node prediction in financial applications and social media~\cite{jin2021anemone,atkinson2021anomaly}, and integration with LLMs~\cite{text_classification_with_graph,chien2021node}.

Ensuring low latency in computing the embeddings of query nodes is key to GNN serving, as GNN models are increasingly used in time-sensitive tasks, including recommendation~\cite{PinSage2018,virinchi2022recommending,lin2022platogl}, fraud detection~\cite{awsfraudgnn,lu2022bright}, and traffic prediction~\cite{derrow2021eta}. At the same time, the underlying graphs are growing in size;  modern industry graph datasets routinely span a few billion nodes and trillions of edges~\cite{bgl_23} and can consume hundreds of TB of memory.
Distributing such large graph datasets and associated feature vectors across multiple machines is thus inevitable these days.

Unfortunately, achieving low-latency GNN serving for large graphs stored across machines is extremely challenging. 
To compute the embeddings of the query nodes, GNN models need the entire $k$-hop neighborhood. The number of $k$-hop neighbors can grow exponentially with $k$, known as the {neighborhood explosion problem}~\cite{graphsaint_20, GraphSage2017, FastGCN2018}.
Serving query nodes can thus require excessive amounts of memory and computation costs due to the size of feature vectors and adjacency matrix in the large neighborhood.

Furthermore, unlike training, we do not know how an incoming query node will be connected to the rest of the graph; no matter how the large graph is partitioned and stored, the $k$-hop neighbors of a query node and their feature vectors are often spread across multiple machines. As a result, creating the query node's ``computation graph'' (\autoref{fig:background1} (right)), by which the node's embedding is computed based on the neighborhood's feature vectors and aggregation paths along the edges, involves fetching the associated data from remote partitions; the resulting massive data volumes (\autoref{subsub:comm_overhead}) cause high communication overheads, worsening serving latency. 

GNN training systems facing similar issues
adopt mitigation techniques based on {\em approximation}, such as reusing historical embeddings~\cite{gnnautoscale_21,sancus_22,freshgnn} or sampling~\cite{DGL, PaGraph2020,bgl_23, P3,leapgnn}. However, we find they are not directly applicable to serving for two reasons: (a) they do not account for the {\em data dependence} between an incoming query node and the existing graph -- i.e., determining how the query node's connectivity impacts the embeddings of the graph's existing nodes. Ignoring such dependence can lead to unacceptably high accuracy loss (\autoref{subsubsec:acc_drop}).
(b) While existing techniques reduce computation (either by reusing node embeddings or computing embeddings for fewer nodes by sampling), we find that the communication involved is still heavy at inference time because of the underlying graph's rich connectivity and the query node's neighborhood being spread across machines.

We present \sys, a GNN serving system that combines data dependency-aware approximation with a new form of computation parallelism to systematically address the overheads, supporting low-latency GNN serving on modern large graphs with minimal accuracy loss.
Existing serving systems either adopt sampling~\cite{quiver2023, PinSage2018,lin2022platogl,helios} (with concomitant high latency or substantial accuracy loss) or apply to smaller graphs~\cite{lambdagrapher, fograph}, limiting their practical applicability (\autoref{sec:related}).

Similar to some training systems~\cite{gnnautoscale_21, sancus_22}, \sys reuses embeddings of existing nodes after training (which we call ``precomputed embeddings (PEs)'') instead of computing the embeddings in the entire $k$-hop neighborhood; this saves network, compute, and memory resources. Because na\"ively reusing PEs ignores data dependence and undermines accuracy,
\sys uses \emph{Selective Recomputation of Precomputed Embeddings} (SRPE)
(\autoref{sec:srpe}). Here, we identify a small number of neighborhood nodes that impact accuracy substantially and
focus on recomputing their embeddings while reusing PEs for other nodes. We prove that this statistically minimizes approximation errors. We develop a practical heuristic to perform recomputation on such nodes within a budget, where the budget trades off latency with accuracy.

SRPE significantly reduces the size of computation graphs, but it still entails substantial communication  (\autoref{subsec:contribution_of_comps}) due to fetching remote PEs and feature vectors.
To mitigate this overhead, we develop \emph{Computation Graph Parallelism} (CGP) (\autoref{sec:cgp}).
Unlike existing GNN systems where a single machine solely creates computation graphs by fetching required data from remote machines, CGP allows each machine to build a partitioned computation graph using data available in local partitions.
CGP then aggregates the partitioned computations using efficient all-to-all collectives and finally applies \emph{custom merge functions} (tailored to the specific type of GNN model) to the aggregations to compute the final outputs.

We implement \sys based on DGL~\cite{DGL,DistDGL2020}, a popular GNN training framework, and evaluate it on representative graph datasets and GNN models.
Our evaluation results show that the combination of SRPE and CGP helps \sys outperform DGL-based full-graph and approximation-driven baseline serving systems by up to 159$\times$ and 10.8$\times$ in latency, respectively, with minimal accuracy loss (\autoref{sec:eval}).

We make the following contributions in this paper:
\begin{packeditemize}
    \item We show that na\"ively applying approximation techniques and reusing computations leads to large errors, and propose SRPE, which statistically minimizes these errors by identifying the parts of computation graphs that need recomputation to avoid accuracy losses (\autoref{sec:srpe}).
    \item We co-design CGP, a new parallelism technique that reduces communication overheads of SRPE with local aggregation by parallelizing the creation and execution of computation graphs for serving various GNN models (\autoref{sec:cgp}).
    \item We show that \sys is able to outperform state-of-the-art techniques and to achieve up to orders of magnitude lower latency with minimal accuracy loss (\autoref{sec:eval}).
\end{packeditemize}

\section{Background}
\label{sec:background}

\begin{figure}
\centering
  \begin{subfigure}{.225\textwidth}
  \centering
    \includegraphics[width=\linewidth]{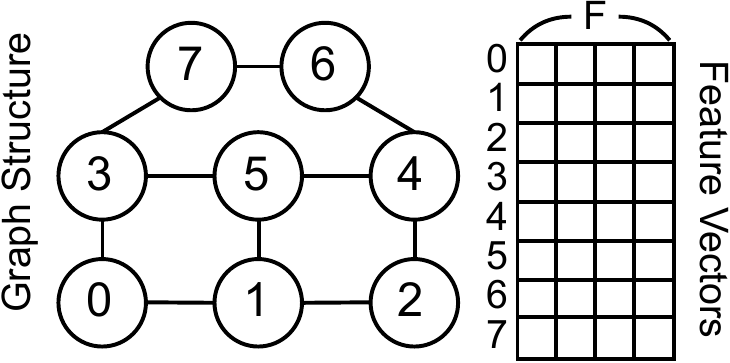}
  \end{subfigure}
  \hspace*{5pt}
  \begin{subfigure}{.225\textwidth}
  \centering
    \includegraphics[width=\linewidth]{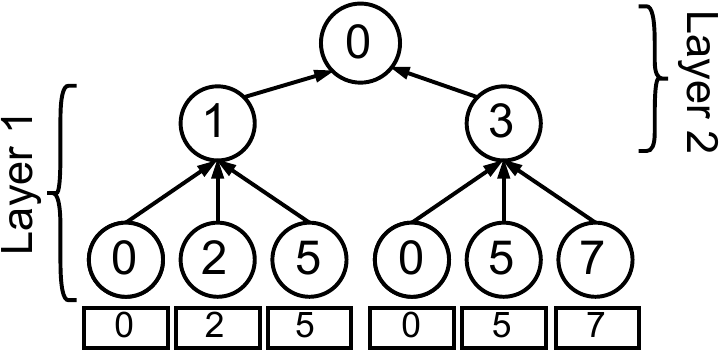}
  \end{subfigure}
  \caption{(\textit{Left}) An example graph dataset with 8 nodes and F-dimensional feature vectors. We use this graph dataset as our running example. (\textit{Right}) The 2-hop computation graph for node 0 where the boxes below represent feature vectors.}
  \label{fig:background1}
\end{figure}

\noindent
{\bf A Primer on GNNs.} Unlike conventional deep neural networks (DNNs) that process independent input vectors without considering their inter-dependencies, GNNs are designed to handle graph datasets. These datasets consist of node feature vectors and a graph structure that expresses the relationships between nodes, as depicted in~\autoref{fig:background1} (left). For instance, in social networks, users are the nodes, their profiles are the feature vectors, and friendships are the edges.

GNN models aggregate the neighborhood information of each node to leverage the relationships and compute more expressive \emph{embeddings} than individual feature vectors alone.
To achieve this, the first step is to create a \emph{computation graph} that contains the $k$-hop neighborhood of each node along with the associated feature vectors, where $k$ is typically 2 or more~\cite{hamilton2017representation}.
A GNN model then \emph{executes} the computation graph to iteratively generate the embedding of a target node.  
Starting with the feature vectors of the farthest $k$-hop neighbors, the GNN model applies DNN operations and an aggregation function at each hop, where each step represents a layer in the GNN model until the embedding for the target node is computed.
\autoref{fig:background1} (right) illustrates an example of a 2-hop computation graph for a node.

The computation of GNN models typically leverages \emph{message passing}~\cite{gilmer2017neural, DGL}, which we can generally express as follows.
For a target node $v$ and layer $1 \leq l \leq k$, define:
\begin{equation} \label{eq:gnn_comp_eq}
    h^{(l)}_v = U^{(l)}(h_v^{(l-1)}, \sideset{}{^{(l)}}{\bigoplus_{u \in \mathit{N}(v)}}\{ M^{(l)}(h^{(l-1)}_u) \})
\end{equation}
 where \smash{$h^{(0)}_u$} is the feature vector for node \smash{$u$}, \smash{$\mathit{N(v)}$} represents direct neighbors of the node \smash{$v$}, and
\smash{$h^{(k)}_v$} is the final embedding for the node \smash{$v$}. Each GNN layer defines \smash{($U$, $\bigoplus$, $M$)} functions, which represent \emph{update}, \emph{aggregate}, and \emph{message} function, respectively.
The message function is applied to the neighbors of node $v$ to create \textit{messages}, which are aggregated by the aggregate function.
The layer embedding of each layer is then obtained by applying the update function to the previous layer embedding of node $v$ and the aggregated messages.

\noindent
{\bf GNN Serving for Large Graphs.}
Similar to conventional DNN models, GNN models can be trained on a static graph dataset in the background.
However, unlike DNN serving, which only needs a trained model, GNN serving additionally requires the graph dataset used in training because of the data dependency in GNN computation.
To serve a large graph dataset (\eg 2 billion nodes and 2 trillion edges~\cite{bgl_23}), the dataset needs to be distributed across multiple machines~\cite{AliGraph2019, DistDGLv2,P3,bgl_23}, and the embeddings for the new {\em query} nodes are computed with this distributed data.

\begin{figure}[t]
 \includegraphics[width=\linewidth]{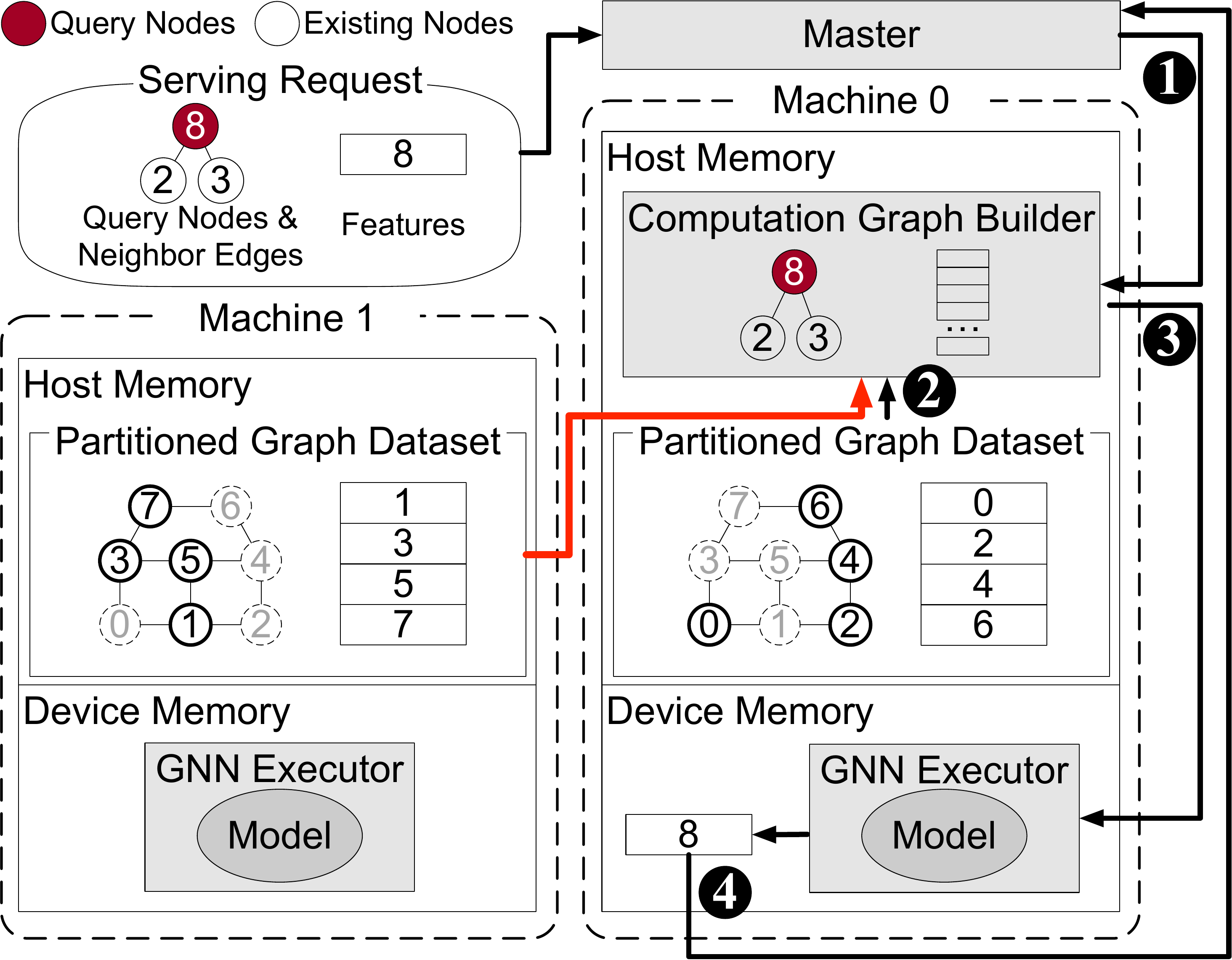}
  \caption{
  Distributed GNN serving end-to-end workflow. To generate the embeddings of (batched) query nodes, \numcircledtikz{1} the master forwards a serving request to one machine. \numcircledtikz{2} A computation graph builder then creates $k$-hop computation graphs for the query nodes by loading from the local partition and fetching required edges and feature vectors from remote partitions. The red line represents remote communication. \numcircledtikz{3} The computation graphs are then executed by a GNN executor after being copied into GPU device memory. \numcircledtikz{4} Finally, the embeddings of the query nodes are returned.}
  \label{fig:gnn_serving_workflow}
\end{figure}

\autoref{fig:gnn_serving_workflow} illustrates a workflow of distributed GNN serving systems.
Here, a training graph dataset is partitioned across multiple machines.
A \emph{serving request} consists of the feature vectors of query nodes and the edges between the query nodes and the existing training nodes. When one of the machines receives the request, it creates computation graphs by fetching the required feature vectors and edges from the other machines, computes the embeddings of the query nodes, and returns the result.

\section{Challenges}
\label{sec:challenges}
Unfortunately, the size of graph datasets that distributed GNN serving systems can handle is limited due to unique properties of modern graph data, especially, the data dependency between query nodes and existing graph nodes, which leads to high overheads. Existing mitigation techniques often fall well short due to fundamental drawbacks. 

\subsection{Overheads from Large Neighborhoods}
\label{subsub:comm_overhead}

\begin{figure}[t]
  \begin{subfigure}{.23\textwidth}
    \includegraphics[width=\linewidth]{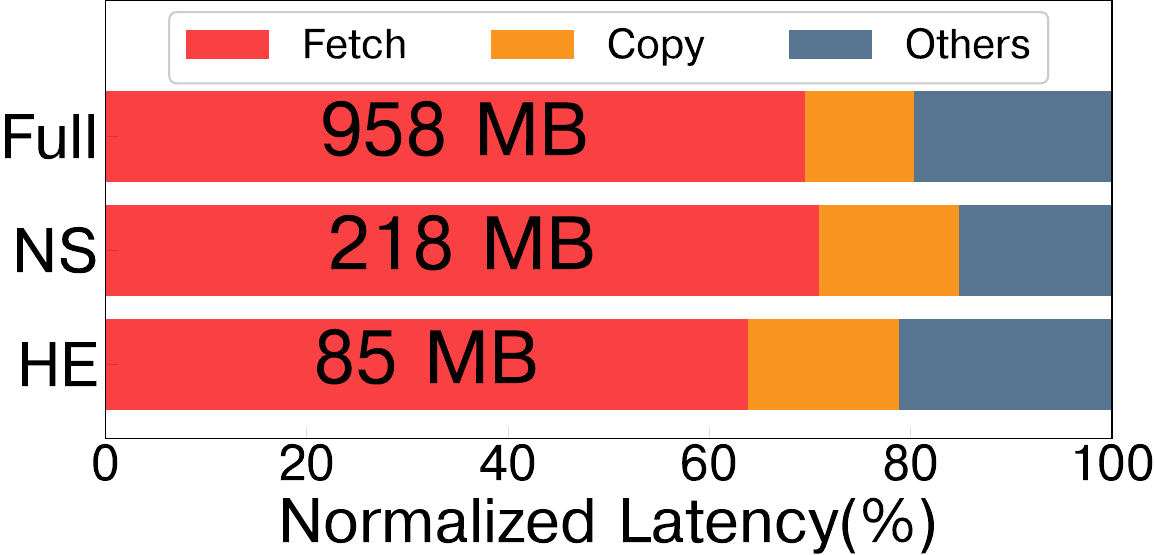}
  \end{subfigure}
  \hspace*{\fill}
  \begin{subfigure}{.23\textwidth}
    \includegraphics[width=\linewidth]{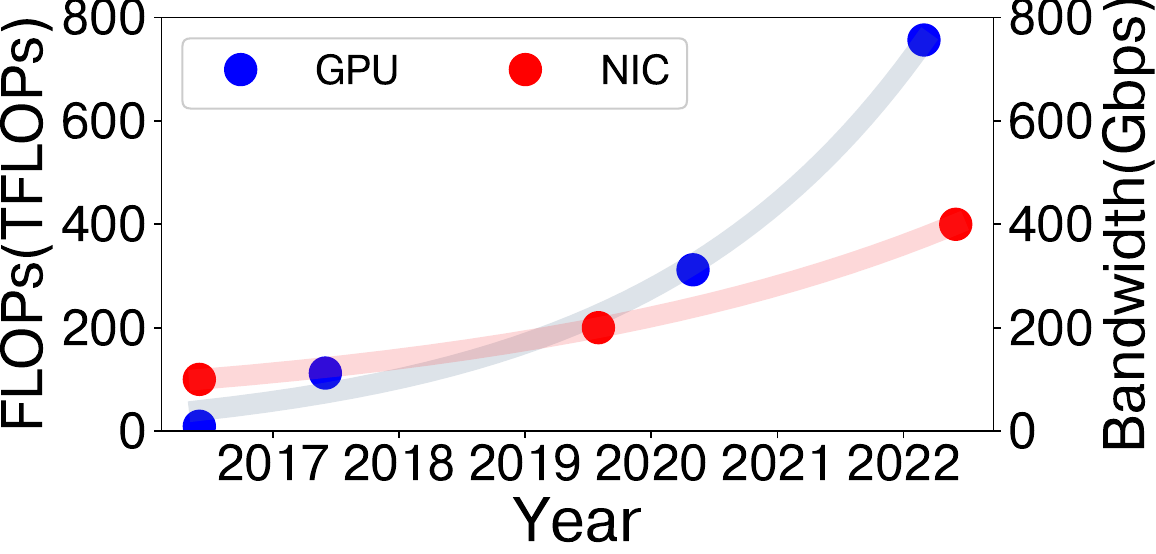}
  \end{subfigure}

  \caption{(Left) Latency breakdown of~\autoref{tab:acc-lat-tradeoff}. The data size in each bar indicates the amount of feature vectors and edges fetched. (Right) Trendlines of peak FP32 FLOPs of NVIDIA GPUs~\cite{nvidia-p100, nvidia-v100, nvidia-a100, nvidia-h100} and bandwidth of NVIDIA NICs~\cite{connectx-4,connectx-5,connectx-6}.
  }
  \label{fig:background}
\end{figure}

GNN serving requires the entire $k$-hop neighborhood in computing the embeddings of query nodes.
Since the number of $k$-hop neighbors can grow exponentially with $k$, memory and compute costs can also be significant to create and execute computation graphs, which contain the feature vectors and edges corresponding to the neighborhood.
For example, serving with the full (neighborhood) computation graph in \autoref{tab:acc-lat-tradeoff} can take 711~ms, likely violating the tight latency SLOs of real-world applications~\cite{click_prediction_1, graph-less, friend_recommendation}.
Here, the computation overhead is substantial, as shown in blue in~\hyperref[fig:background]{Fig.~\ref*{fig:background}}.
Also, the memory overhead from these large neighborhoods can easily lead to GPU out-of-memory errors despite their tens of GB memory capacity (\autoref{sec:overall_perf}).

Moreover, it is hard to predict how query nodes are connected to the rest of the graph and as such feature vectors and edges can be stored across multiple machines (as shown in \autoref{fig:gnn_serving_workflow}); thus, creating computation graphs involves fetching neighborhood information and the corresponding feature vectors from remote partitions (Step \numcircledtikz{2}), resulting in high communication overhead, shown in red in \hyperref[fig:background]{Fig.~\ref*{fig:background}}.
\begin{table}
    \small 

    \centering
    \begin{tabular}{p{3.95cm}|p{1.65cm}|p{1.7cm}}
        Method & Latency (ms) & Accuracy (\%) \\
    \midrule
        Full Computation Graph (Full) & 711.1 & 56.9 \\
        Neighborhood Sampling (NS) & 168.8 & 50.9 (-6.0) \\
        Historical Embeddings (HE) & 66.4 & 50.6 (-6.3) \\
    \end{tabular}
    \caption{Latency and accuracy of different serving methods. We run serving requests of batch size 1,024 with 4 machines. A 3-Layer Graph Attention Networks~\cite{GAT2018} trained on the Yelp dataset~\cite{graphsaint_20} is used. We describe the detailed setup in~\autoref{subsec:exp_setup}.}
    \label{tab:acc-lat-tradeoff}
\end{table}

\subsection{Limitations of Today's Approximations}
\label{subsubsec:acc_drop}

To mitigate problems due to large neighborhoods and overheads, one could employ approximation taking inspiration from similar techniques in GNN {\em training} systems. Unfortunately, this can lead to significant model accuracy loss~(\autoref{tab:acc-lat-tradeoff}) and, in some cases, may not improve latency.

\noindent
{\bf Historical Embeddings.}
Instead of creating full computation graphs every iteration, GNN training systems~\cite{gnnautoscale_21, sancus_22, freshgnn} reuse the layer embeddings from previous iterations, called \emph{historical embeddings}~\cite{gnnautoscale_21,vr_gcn_18}.
This technique avoids the exponential expansion of computation graphs, allowing GNN training systems to construct computation graphs only with direct neighbors.
Since model parameters are updated at each iteration, historical embeddings introduce approximation errors.
Over multiple training epochs, the training systems address the errors using an auxiliary loss function or updating historical embeddings based on their staleness~\cite{gnnautoscale_21,sancus_22,freshgnn}.
One may consider applying this technique to serving by reusing the layer embeddings from the last training epoch.
However, as the embeddings are data dependency-unaware, \ie they do not account for how new query nodes impact existing embeddings, using them for serving can hurt accuracy. As shown in~\autoref{tab:acc-lat-tradeoff}, although serving can be 10.7$\times$ faster with this technique, it results in a 6.3\% points drop in accuracy.

\noindent
{\bf Sampling.}
GNN training systems~\cite{PyG2019, DGL, PaGraph2020, gnnlab_22, bgl_23, legion_23, P3, AGL2020, AliGraph2019, DistDGL2020,leapgnn} employ \emph{sampling} with which they aggregate only a small number of sampled neighbors at each hop~\cite{graphsaint_20,GraphSage2017,FastGCN2018,ClusterGCN2019}.
This drastically reduces the size of the computation graph at the cost of accuracy due to approximation errors from sampling.
Multiple training epochs are executed to account for the accuracy loss, and each epoch samples the computation graph differently. Running many epochs leads to convergence.
However, in serving, we find that a sampled computation graph can hurt accuracy significantly (6.0\% points drop in \autoref{tab:acc-lat-tradeoff}) since we do not have multiple chances to recover approximation errors from sampling.
Moreover, latency can be still high because, even with sampling, the size of computation graphs and the associated communication overhead can be significant (\autoref{sec:overall_perf}).

Furthermore, even with either approximation, over 80\% of the latency is still spent fetching the associated feature vectors (\hyperref[fig:background]{Fig.~\ref*{fig:background}}), where feature dimensions typically range from hundreds to thousands~\cite{ogb-lsc,PinSage2018,AGL2020}, from remote partitions and copying them into the GPU memory.
We observe that the communication overhead is a key bottleneck in GNN serving for large graphs regardless of GNN models and graph datasets (\autoref{subsec:contribution_of_comps}). Recent trends in GPU and NIC performance~(\autoref{fig:background} (right)) further indicate that communication will remain a challenge in large-scale GNN serving.

\begin{figure}
  \begin{subfigure}{.225\textwidth}
    \includegraphics[width=\linewidth]{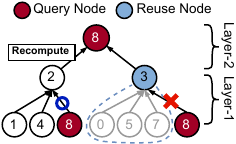}
  \end{subfigure}
  \hspace*{5pt}
  \begin{subfigure}{.225\textwidth}
    \includegraphics[width=\linewidth]{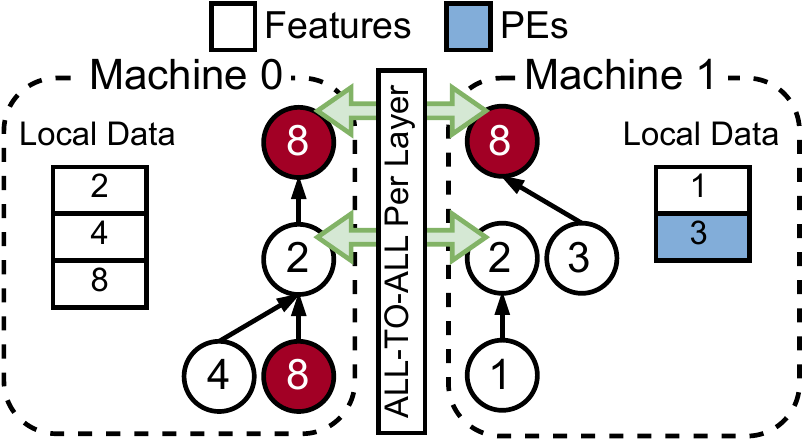}
  \end{subfigure}
  \caption{High-level illustration of (\textit{Left}) Selective Recomputation of Precomputed Embeddings (SRPE) and (\textit{Right}) Computation Graph Parallelism (CGP). The query node 8 is connected to the existing nodes 2 and 3 of the example graph dataset in~\autoref{fig:background1}.
}
  \label{fig:motiv1}
\end{figure}

\section{\sys Overview}\label{sec:overview}

Our system, \sys, addresses GNN serving challenges and overheads due to large neighborhoods through two complementary techniques: smart data dependency-aware approximation and distributed computation graph creation and execution.

\noindent
{\bf T-1: Selective Recomputation of Precomputed Embeddings (\autoref{sec:srpe}).}
To deal with neighborhood explosion, we design a technique called selective recomputation of precomputed embeddings (SRPE) that reduces redundant computations of layer embeddings by effectively reusing previously computed embeddings named \emph{precomputed embeddings} (PE). Unlike historical embeddings used in training 
 (\autoref{subsubsec:acc_drop}), SRPE mitigates accuracy drops by \emph{selectively recomputing} a portion of the PEs that contributes to high approximation errors.
SRPE decides on recomputation nodes based on a simple but effective heuristic policy that statistically minimizes the approximation errors.
\autoref{fig:motiv1} (left) illustrates an example usage of SRPE.
After training, SRPE precomputes the embedding of existing training nodes.
When computing the embedding of query node 8, SRPE reuses the PE of node 3 while \emph{recomputing} the PE of node 2 to better reflect the contribution of node 8 (\ie the edge from node 8 to 2 at Layer-1).

\begin{figure}[t]
 \includegraphics[width=\linewidth]{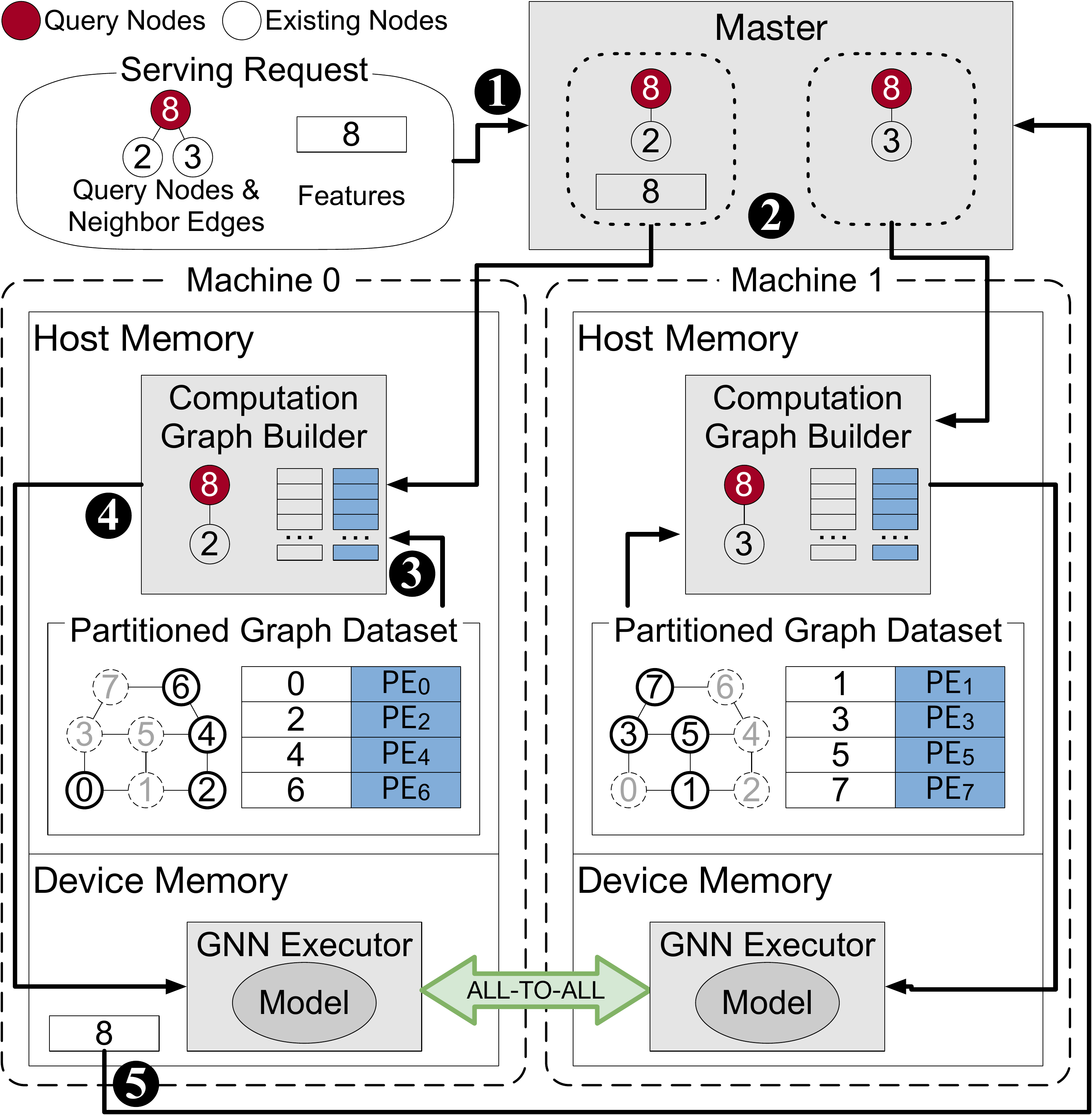}
  \caption{\sys end-to-end workflow.}
  \label{fig:omega_overview}
\end{figure}

\noindent
{\bf T-2: Computation Graph Parallelism and Custom Merge Functions (\autoref{sec:cgp}).}
While SRPE reduces the size of computation graphs and corresponding computational overheads, it still has high 
communication overhead due to fetching PEs and feature vectors; \eg we find that SRPE alone can still yield latencies as high as 978~ms (\autoref{subsec:contribution_of_comps}) due to the overhead.
To further minimize this overhead, we develop \emph{computation graph parallelism} (CGP), which distributes both \emph{creation} and \emph{execution} of computation graphs across multiple machines.
In CGP, each machine carefully constructs a \emph{partitioned computation graph} to use only local feature vectors and PEs during execution, thus avoiding heavy remote communication.
For instance, in \autoref{fig:motiv1} (right), to compute the Layer-1 embedding of node 2, Machine 0 uses local feature vectors of nodes 4 and 8, while Machine 1 uses the local feature vector of node 1.
However, since each machine has a partial view of the entire neighborhood, conventional message passing computation (\autoref{eq:gnn_comp_eq}) fails to generate correct outputs.
To overcome this challenge, CGP employs a form of distributed execution that extends message passing with collective communications and \emph{custom merge functions} tailored to various GNN models.

\noindent
{\bf End-to-end Workflow.}
\autoref{fig:omega_overview} illustrates the end-to-end workflow of \sys.
The PEs of all nodes in the training graph dataset are computed after training and distributed across machines along with the graph dataset.
\numcircledtikz{1} Upon the arrival of a serving request containing a batch of query nodes, the master server evenly splits the query nodes, their feature vectors, and the neighbor edges.
\numcircledtikz{2} The partitioned requests are then distributed to the computation graph builder in each machine, 
\numcircledtikz{3} which creates a partitioned computation graph.
When building the computation graph, each builder determines which PEs to {\em recompute} and adds the edges required for the recomputation to the computation graph.
\numcircledtikz{4} Copied to GPU memory, the partitioned computation graphs are executed by GNN executors, which employ local aggregations followed by collective communications to generate the final output embeddings with customized merge functions for GNN models; \numcircledtikz{5} the master server collects and returns the output embeddings.

\noindent
{\bf Problem Scope.}
This paper focuses on achieving low-latency, large-scale GNN serving for \textit{new} query nodes that are unseen during training, while mitigating large computational overheads and approximation errors (\autoref{sec:challenges}).
We assume that GNN models are trained on a fixed graph (\eg the example graph dataset in~\autoref{fig:background1}), and during the serving phase, query nodes arrive with edges connecting them to training nodes in the fixed graph (\eg query node 8 in~\autoref{fig:omega_overview} is connected to nodes 2 and 3).
Addressing potential staleness problems and enabling dynamic updates of graph datasets, GNN models, and PEs is left for future work.

\section{Selective Recomputation of PE}\label{sec:srpe}
To mitigate the overheads from large neighborhoods (\autoref{sec:challenges}), \sys reuses precomputed embeddings (PEs), \smash{$p_u^{(l)}$} for each node $u$ and layer $l$ (\smash{$1 \leq l \leq k-1$}), which are captured from the layer embeddings (\ie \smash{$h_u^{(l)}$} in \autoref{eq:gnn_comp_eq}) after training. During the serving phase, for a query node $v$, \sys computes the final embedding \smash{$h_v^{(k)}$} without fully recalculating the embeddings for the full $k$-hop neighborhood. Instead, it substitutes the embeddings of $v$'s direct neighbors with their respective PEs, utilizing \smash{$p_u^{(l)}$} in place of computing \smash{$h_u^{(l)}$}. This approach significantly reduces both computational complexity and memory requirements from \smash{$O(N^k)$} to \smash{$O(N \times k)$}, where \smash{$N$} represents the average number of neighbors.

\subsection{Skewed Approximation Errors of PEs}\label{subsec:skewed_errors}
However, we observe that na\"ively reusing PEs for GNN serving can lead to substantial reductions in model accuracy, such as 6.3\% points decrease (denoted HE in \autoref{tab:acc-lat-tradeoff}).
This reduction is attributed to approximation errors in PEs. These errors occur because PEs are snapshotted after training, failing to aggregate the embeddings from new query nodes.
For example, the PE of node 3 shown in \autoref{fig:motiv1} (left) is computed considering only the neighbors in the training graph (nodes 0, 5, 7), excluding the new query node 8.

We quantify the approximation errors by computing the difference between the \textit{full embeddings}, \smash{$f_u^{(l)}$}, which include all the edges from query nodes in aggregation, and the PEs.
When $u$ is a direct neighbor of query nodes, we compute its PE approximation error as \smash{$\sum_{l=1}^{k-1}||f_u^{(l)} - p_u^{(l)}||$}.

\begin{figure}
  \begin{subfigure}{.225\textwidth}
    \includegraphics[width=\linewidth]{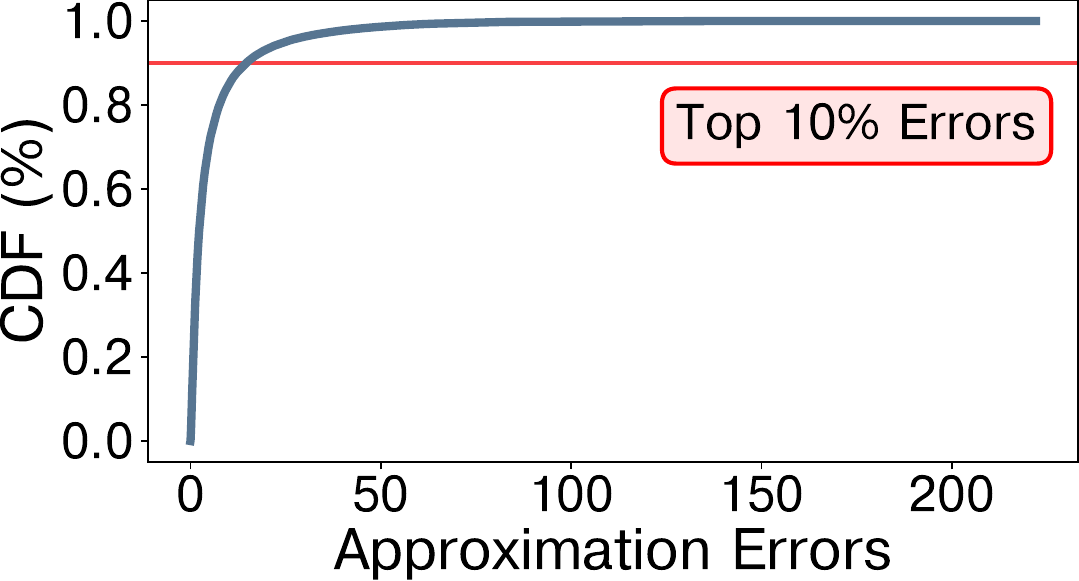}
  \end{subfigure}
  \hspace*{5pt}
  \begin{subfigure}{.225\textwidth}
    \includegraphics[width=\linewidth]{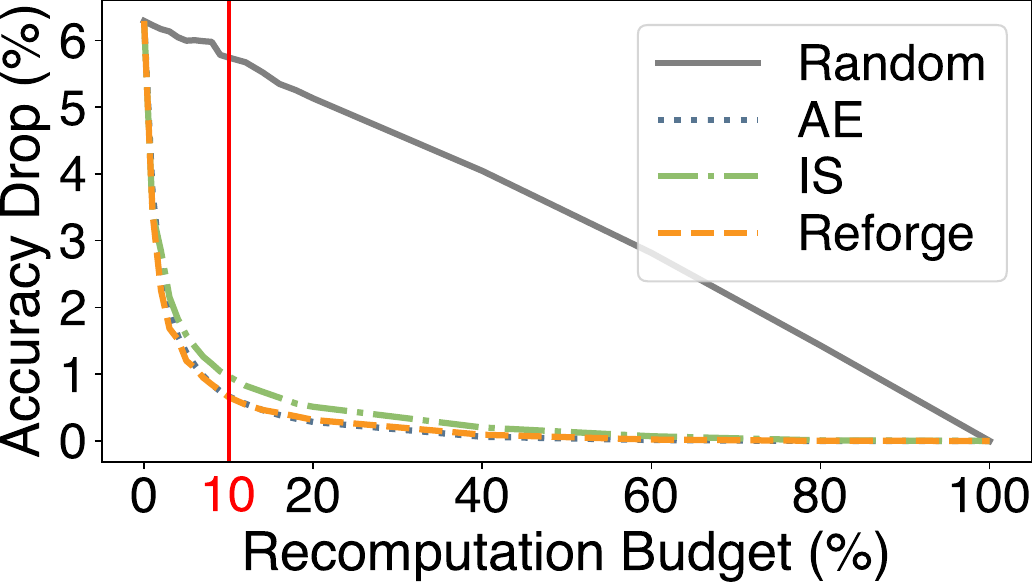}
  \end{subfigure}  
  \caption{(\textit{Left}) CDF of the approximation errors of PEs, derived from the workload in \autoref{tab:acc-lat-tradeoff}. We randomly selected 25\% of test nodes as query nodes, computed PEs with the remaining nodes, and aggregated the errors using the query nodes. (\textit{Right}) Effectiveness of various recomputation policies in restoring accuracy with increasing recomputation budgets. \textsf{AE} and \textsf{IS} represent recomputation based on actual approximation errors and node importance scores. \sys denotes the proposed query edge ratio-based policy (\autoref{subsubsec:policy}).}
  \label{fig:srpe_error_analysis}
\end{figure}

We observe that the approximation errors are {\em highly skewed} -- the errors from a small number of PEs mainly contribute to accuracy losses.
\autoref{fig:srpe_error_analysis} (left) shows a distribution of the approximation errors of PEs.
The top 10\% of PE approximation errors are several orders larger than the other 90\%.
More importantly, recomputing the 10\% of PEs, while reusing the other PEs, effectively recovers accuracy drop from -6.3\% points to -0.7\% points (\textsf{AE} in \autoref{fig:srpe_error_analysis} (right)). Exploiting the skew in approximation errors is key to the effectiveness of recomputation; e.g., randomly selecting the 10\% recomputation targets (\textsf{Random} in \autoref{fig:srpe_error_analysis} (right)) yields a marginal accuracy benefit (from -6.3\% points to -5.7\% points).

\subsection{Policy for Selective PE Recomputation}\label{subsec:recom_policy}
Based on the observation, we propose to \textit{selectively} recompute PEs that show high approximation errors when used to serve query nodes. The key challenge in this approach lies in devising an effective \textit{recomputation policy} that can identify PEs prone to high approximation errors.

\subsubsection{Problem Formulation}
\label{subsubsec:problem_formulation}
\begin{figure}
\centering
  \begin{subfigure}{.225\textwidth}
  \centering
    \includegraphics[width=\linewidth]{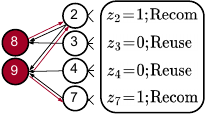}
  \end{subfigure}
  \hspace*{5pt}
  \begin{subfigure}{.225\textwidth}
  \centering
    \includegraphics[width=\linewidth]{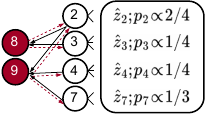}
  \end{subfigure}
  \caption{
An example of selective recomputation in a batch of two query nodes: 8 (connected to nodes 2 and 3) and 9 (connected to nodes 2, 4, and 7).
(\textit{Left}) Recomputation of two candidates given a 50\% recomputation budget. (\textit{Right}) Random variables with recomputation probabilities such that $\sum p_i = 0.5$. \sys's policy assigns each probability proportional to the ratio of edges from the query nodes.
}
  \label{fig:srpe_recom}
\end{figure}

We describe the problem of selective PE recomputation and the objective of recomputation policies, using an example of serving two query nodes (\ie nodes 8 and 9) in \autoref{fig:srpe_recom} (left).
In this example, we can utilize PEs of the direct neighbors of the query nodes (\ie nodes 2, 3, 4, and 7), which we call \textit{recomputation candidates}.
A policy decides whether to recompute or reuse the PE for each candidate node.
Recomputation includes the edges from query nodes to a candidate node, whereas reuse ignores them.
Then, the objective of the recomputation policy is to find the best recomputation targets, given a recomputation budget, that minimizes the approximation errors of the candidates.
\autoref{fig:srpe_recom} (left) illustrates an example where the budget allows for two out of four PEs to be recomputed.

The recomputation policy can be formulated as a constrained optimization problem as follows:
\begin{equation}
\begin{aligned}
    \min_{z \in \{ 0, 1 \}^{|\mathit{R}|}} \sum_{u \in \mathit{R}}\sum_{l=1}^{k-1}||f_u^{(l)} - h_{G_z, u}^{(l)}||
    \:\:\:\textrm{\textit{sub. to.}}\:\: \sum_{u \in \mathit{R}} z_u \leq \gamma|\mathit{R}|
\end{aligned}
\end{equation}
Here, $\mathit{R}$ is the recomputation candidates, $z$ is a $|R|$-sized 0-1 vector deciding recomputation of each candidate, $G_z$ denotes the corresponding graph, $h_{G_z, u}$ represents the approximated embeddings in $G_z$, and $f_u$ are the full embeddings. Finally, $\gamma$ refers to the budget.

However, directly solving the problem at serving time is challenging because it is impractical to compute the full embeddings.
Instead, we stochastically approximate the constrained optimization problem.
As illustrated in \autoref{fig:srpe_recom} (right), we assign independent binary random variables $\hat{z}_u$ to each candidate $u$ such that it has a recomputation probability of $p_u=\mathbb{E}[\hat{z}_u]$, within the budget $\gamma = \sum_{u} p_u$.

In this setting, our goal is to find the optimal probabilities that minimize the approximation errors.
To achieve this, we adopt the variance minimization technique from recent work in sampling-based GNN training~\cite{graphsaint_20}; we define unbiased estimators for the full embeddings and derive the optimal recomputation probabilities that minimize the variance of the estimators.
By selecting candidates with high probabilities, we can indirectly reduce the approximation errors.

We develop unbiased estimators for the full embeddings.
For each $u$, the full embedding \smash{$f_u^{(l)}$} is computed with aggregations from the query nodes (\smash{$q_u^{(l)}$}) and those from the training nodes (\smash{$t_u^{(l)}$}).
Here, we simplify the aggregation as the mean of the messages from neighbors; for example, \smash{$q_u^{(l)} = \sum_{v \in \mathit{N_Q}(u)} m_v^{(l)}/|\mathit{N}(u)|$} where \smash{$\mathit{N_Q}(u)$} denotes the query nodes connected to $u$ and $m_v^{(l)}$ are the messages.
Then, we define the estimator \smash{$\hat{f}_u^{(l)} = \frac{1}{p_u}\hat{z}_u q_u^{(l)} + t_u^{(l)}$} such that $\mathbb{E}[\hat{f}_u^{(l)}] = f_u^{(l)}$.
We obtain the optimal probabilities that minimize the variance of the estimators with the following theorem.
\begin{theorem}
\label{theorem:minvar}
The sum of the variances of every dimension of the estimators ($\sum_u \sum_{l=1}^{k-1} \hat{f}_u^{(l)}$) is minimized when $p_u \propto ||\sum_{l=1}^{k-1} q_u^{(l)}|| = ||\sum_{l=1}^{k-1}\sum_{v \in \mathit{N_Q}(u)} \frac{m_v^{(l)}}{|\mathit{N}(u)|}||$, given $\gamma = \sum_{u} p_u$.
\end{theorem}
The proof is based on \textit{Theorem 3.2} from \cite{graphsaint_20}. With the constant sum of probabilities $\gamma$, we can derive the lower bound of the estimator using the Cauchy-Schwarz inequality. Equality holds when recomputation probabilities align with those specified in the theorem. We provide the formal proof of \autoref{theorem:minvar} in Appendix~\ref{appendix:proof}.

\subsubsection{The Top-Query-Edges-Ratio Policy}
\label{subsubsec:policy}
Based on the analysis, a recomputation policy can calculate the optimal probabilities and recompute the PEs within the top $\gamma$\% of these probabilities.
However, the exact calculation requires the full embeddings of query nodes to obtain the messages ($m_v^{(l)}$) of query nodes, which is infeasible in serving.
Thus, instead of computing the full embeddings, we adopt the simplification from prior work~\cite{graphsaint_20}, whereby our recomputation policy approximates $p_u \propto \frac{|\mathit{N}_Q(u)|}{|\mathit{N}(u)|}$ without considering the message terms.
We explain the intuition and highlight the policy's effectiveness below.

We name our policy \textit{top-query-edges-ratio} since it recomputes PEs with higher ratios of edges from query nodes.
For instance, in \autoref{fig:srpe_recom} (right), the policy recomputes the PEs of nodes 2 and 7.
Intuitively, a higher ratio of query edges is likely to change the PE significantly and result in a large approximation error.
Empirically, \autoref{fig:srpe_error_analysis} (right) shows this policy (\sys) can recover accuracy drops almost similarly to recomputation based on real approximation errors (\textsf{AE}).

To further assess the effectiveness of \sys's policy, we compare it with two alternatives, \textsf{IS} and \textsf{Random}, as shown in \autoref{fig:srpe_error_analysis} (right). \textsf{IS} chooses recomputation targets based on node importance scores, defined for each node $v$ as $IS(v) = \frac{1}{\text{deg}(v)}\sum_{u \in N(v)}\frac{1}{\text{deg}(u)}$, following existing sampling-based training methods \cite{graphsaint_20,FastGCN2018,vr_gcn_18}. 
Though beneficial for reducing sampling variance during training, these scores focus more on existing nodes and are less effective for recomputation during serving, resulting in less optimal performance compared to \sys's policy.
\textsf{Random} selects targets randomly. This approach fails to identify error-prone PEs, leading to ineffective recovery of accuracy losses, even with large recomputation budgets.

In \autoref{tab:srpe_table}, we evaluate \sys's policy across representative GNN models and graph datasets, which shows it can effectively recover accuracy to within 1\% point with minimal budget.
Our policy consistently outperforms the alternatives, \textsf{IS} and \textsf{Random}, similar to the result in \autoref{fig:srpe_error_analysis} (right).
We extensively evaluate the recomputation policies across diverse GNN models and graph datasets, and observe consistent results, as detailed in Appendix~\ref{appendix:srpe_eval}.

Finally, \sys allows for changing the amount of recomputation through the recomputation budget parameter $\gamma$ to navigate the tradeoff between accuracy and latency.
Users can adjust the parameter depending on the model, dataset, and acceptable percentage of accuracy drop and latency.
We evaluate this tradeoff in~\autoref{subsec:overhead_of_pe} and show that \sys can recover accuracy drops with minimal latency increase. We plan to investigate automatic selection of the recomputation budget based on runtime distributions of query nodes in future work.

\section{Computation Graph Parallelism}\label{sec:cgp}

While SRPE mitigates the neighborhood explosion problem, communication overhead remains significant.
We observe that over 80\% of the latency is due to creating computation graphs, which involves fetching feature vectors, PEs, and neighbor edges from remote machines and copying them to GPU memory (\autoref{subsec:contribution_of_comps}).
We now describe \emph{computation graph parallelism} (CGP) to further reduce communication overhead by distributing computation graph creation and execution.

\begin{figure}
\centering
  \begin{subfigure}{.225\textwidth}
  \centering
    \includegraphics[width=\linewidth]{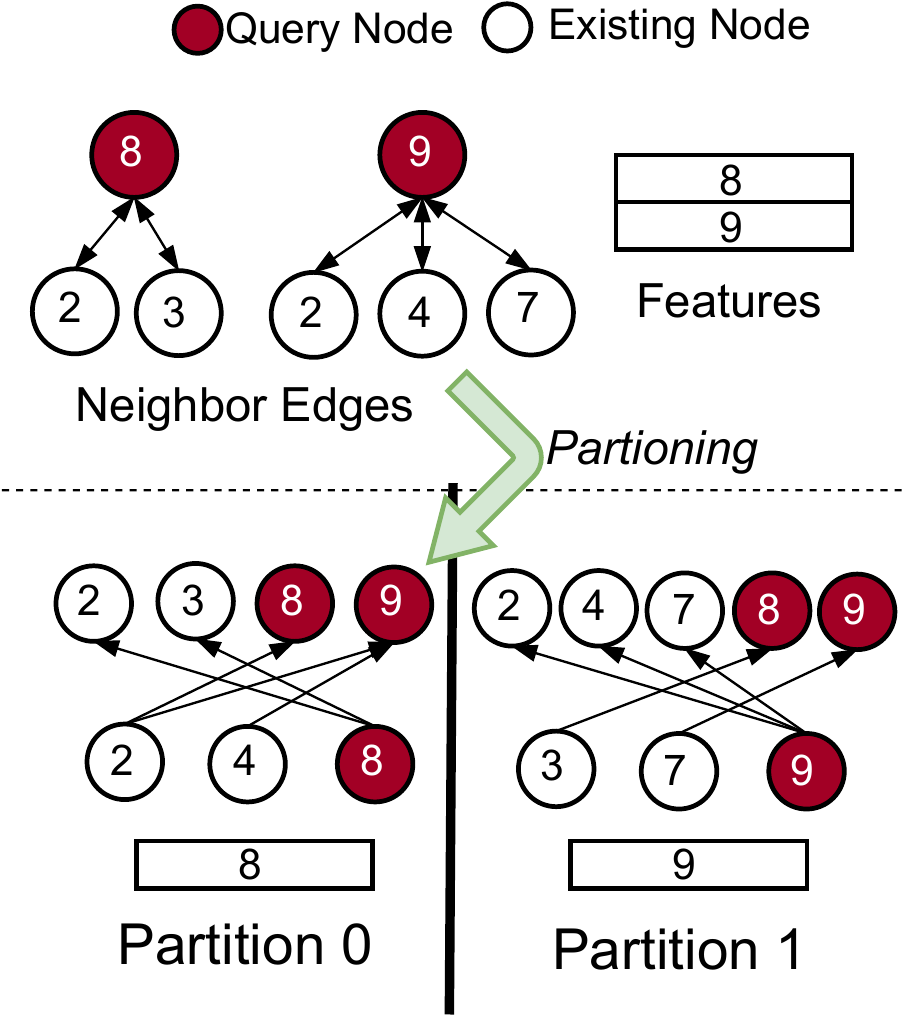}
    \caption{Serving Request}
    \label{fig:cgp_1a}
  \end{subfigure}
  \hspace*{5pt}
  \begin{subfigure}{.225\textwidth}
  \centering
    \includegraphics[width=\linewidth]{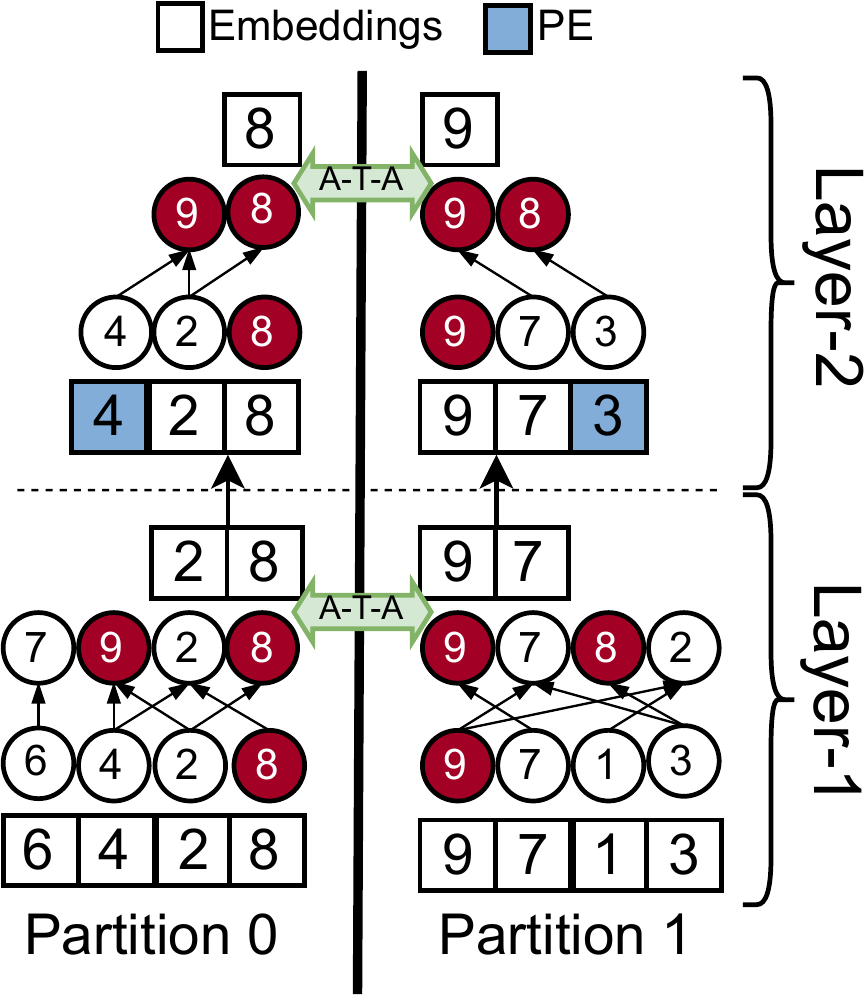}
    \caption{Computation Graph}
    \label{fig:cgp_1b}
  \end{subfigure}
  \caption{
  (a) An example serving request on the graph dataset in~\autoref{fig:background1} with the partitioned requests depicted at the bottom. (b) Partitioned computation graphs for the request.
  }
  \label{fig:cgp_1}
\end{figure}

\begin{algorithm}[t]
\scriptsize
\caption{\sys Distributed Execution of CGP}
\label{algo:cgp}
\begin{algorithmic}[1]
\Statex \textbf{Params:} $n\_parts$: number of partitions, $\gamma$: recomputation ratio, $n\_layers$: number of GNN layers, $q\_n$: query nodes, $q\_e$: query edges, $q\_f$: query features
\vspace{4pt}
\State \textbf{def }{Serving\_Request\_Partitioning}{$(q\_n, q\_e, q\_f)$:}
    \State \quad $part\_reqs \gets [\;]$
    \State \quad \textbf{for } $p\_i \gets 0 \text{ to } n\_parts-1$ \textbf{ do}
        \State \quad \quad $p\_e \gets \{ ( u \rightarrow v ) \in q\_e | part(u) = p\_i\}$
        \State \quad \quad $p\_f \gets \{ q\_f[w] | w \in q\_n, part(w) =  p\_i\}$
        \State \quad \quad $part\_reqs \gets part\_reqs\cup \{q\_n, p\_e, p\_f\}$
    \State \quad \Return $part\_reqs$
\vspace{6pt}
\State \textbf{def }{Computation\_Graph\_Generation}{$(q\_n, p\_e)$:}
    \State \quad $last\_cg \gets ( q\_n, p\_e )$ \Comment{Destination nodes and edges}
    \State \quad $rc\_cands \gets srcs(last\_cg) \backslash q\_n$
    \State \quad $rc\_n \gets$ \texttt{\;Nodes with top $\gamma$\;\% query-edge-ratio in $rc\_cands$}
    \State \quad $reuse\_n \gets rc\_cands \backslash rc\_n$
    \State \quad $rc\_n \gets ALL\_GATHER(rc\_n)$
    \State \quad $rc\_e \gets local\_part.edges(rc\_n)$
    \State \quad $rc\_cg \gets ( q\_n \cup rc\_n, p\_e \cup rc\_e )$
    \State \quad \Return $last\_cg, rc\_cg, reuse\_n$
\vspace{6pt}
\State \textbf{def }{Computation\_Graph\_Execution}{$(last\_cg, rc\_cg, reuse\_n, p\_f)$:}
    \State \quad $n \gets srcs(rc\_cg) \backslash q\_n$
    \State \quad $embs \gets local\_part.feats(n) \cup p\_f$
    \State \quad \textbf{for } $l \gets 1 \text{ to } n\_layers - 1$ \textbf{ do}
        \State \quad \quad $embs \gets Dist\_Layer\_Exec(rc\_cg, embs)$
        \State \quad \quad \textbf{if } $l < n\_layers - 1$ \textbf{ then}
            \State \quad \quad \quad $n \gets srcs(rc\_cg) \backslash dsts(rc\_cg)$
            \State \quad \quad \quad $embs \gets embs \cup local\_part.pes(n, l)$
        \State \quad \quad \textbf{else}
            \State \quad \quad \quad $embs \gets embs \cup local\_part.pes(reuse\_n, l)$
    \State \quad \Return $Dist\_Layer\_Exec(last\_cg, embs)$
\end{algorithmic}
\end{algorithm}

\subsection{Distributed Creation and Execution}

\label{subsec:cgp_creation}

We describe the key steps in CGP (\autoref{algo:cgp}) using an example serving request with two batched query nodes (\ie nodes 8 and 9) at the top of \autoref{fig:cgp_1a}.

\noindent
{\bf Serving Requests Partitioning.}
As described in~\autoref{sec:overview}, \sys's master first splits incoming serving requests. Specifically, the master evenly assigns partitions to every query node in a request. In our example, nodes 8 and 9 are assigned to partitions 0 and 1, respectively. The edges in a request are then split based on the source nodes' partitions to enable local aggregation. For instance, the edges whose source nodes are 2, 4, and 8 are included in the partitioned request for partition 0, as depicted at the bottom left of \autoref{fig:cgp_1a}.
The feature vectors of query nodes follow the partitions of the query nodes.
The master sends the partitioned edges and features to corresponding computation graph builders.

\noindent
{\bf Computation Graph Generation.}
Among the edges, each builder uses those having query nodes as destination nodes to create the last layer of a computation graph (\eg Layer-2 in~\autoref{fig:cgp_1b}).
Each builder then applies \sys's recomputation policy (\autoref{subsubsec:policy}) to identify recomputation target nodes among the input nodes of the last layer. In Layer-2 of~\autoref{fig:cgp_1b}, while the PEs of nodes 3 and 4 are reused, the embeddings of nodes 2 and 7 are recomputed.
For the recomputation, all of the edges terminating at the target nodes need to be included in the other layers. Since the edges are distributed across different partitions, the builders first broadcast their recomputation target nodes to each other through all-gather collective (\eg nodes 2 and 7), extract the required edges from local graphs (\eg $4 \shortrightarrow 2$ and $6 \shortrightarrow 7$ in Layer-1 of partition 0 in~\autoref{fig:cgp_1b}) and from the partitioned requests (\eg $8 \shortrightarrow 2$), then create the remaining layers (\eg Layer-1 in~\autoref{fig:cgp_1b}).

\noindent
{\bf Layer-wise Distributed Execution.}
With the computation graphs, GNN executors compute the embeddings of the query nodes by executing GNN layers sequentially.
To first compute Layer-1 embeddings, each executor computes partial aggregations using its local features, shuffles them with the other executors through collective communications, and merges them.
The output embeddings are concatenated with the local PEs, which are used as the input for the next layer (\eg Layer-2 input embeddings in ~\autoref{fig:cgp_1}).
The process is repeated until the query nodes' embeddings are computed in the last layer execution.

To enable the distributed execution, \sys extends conventional message passing (\autoref{eq:gnn_comp_eq}).
Instead of applying the aggregation function to the entire neighborhood, \sys generates a local aggregation for each partition using the neighbors placed in each partition.
\sys further defines a merge function to compute global aggregations with the local aggregations. The $l$-th layer execution is as follows:
\begin{equation} \label{eq:gnn_comp_eq_2}
\begin{aligned}
    a^{(l)}_{v, p} &= \hat{\bigoplus}^{(l)}_{u \in \mathit{N}_p(v)}\{ M^{(l)}(h^{(l-1)}_u) \}, 0 \leq p < P \\
    h^{(l)}_v &= U^{(l)}(h_v^{(l-1)}, \sideset{}{^{(l)}}{\biguplus }\{ a^{(l)}_{v, p} \}_{0 \leq p < P})
\end{aligned}
\end{equation}

\begin{figure}
\centering
 \includegraphics[width=\linewidth]{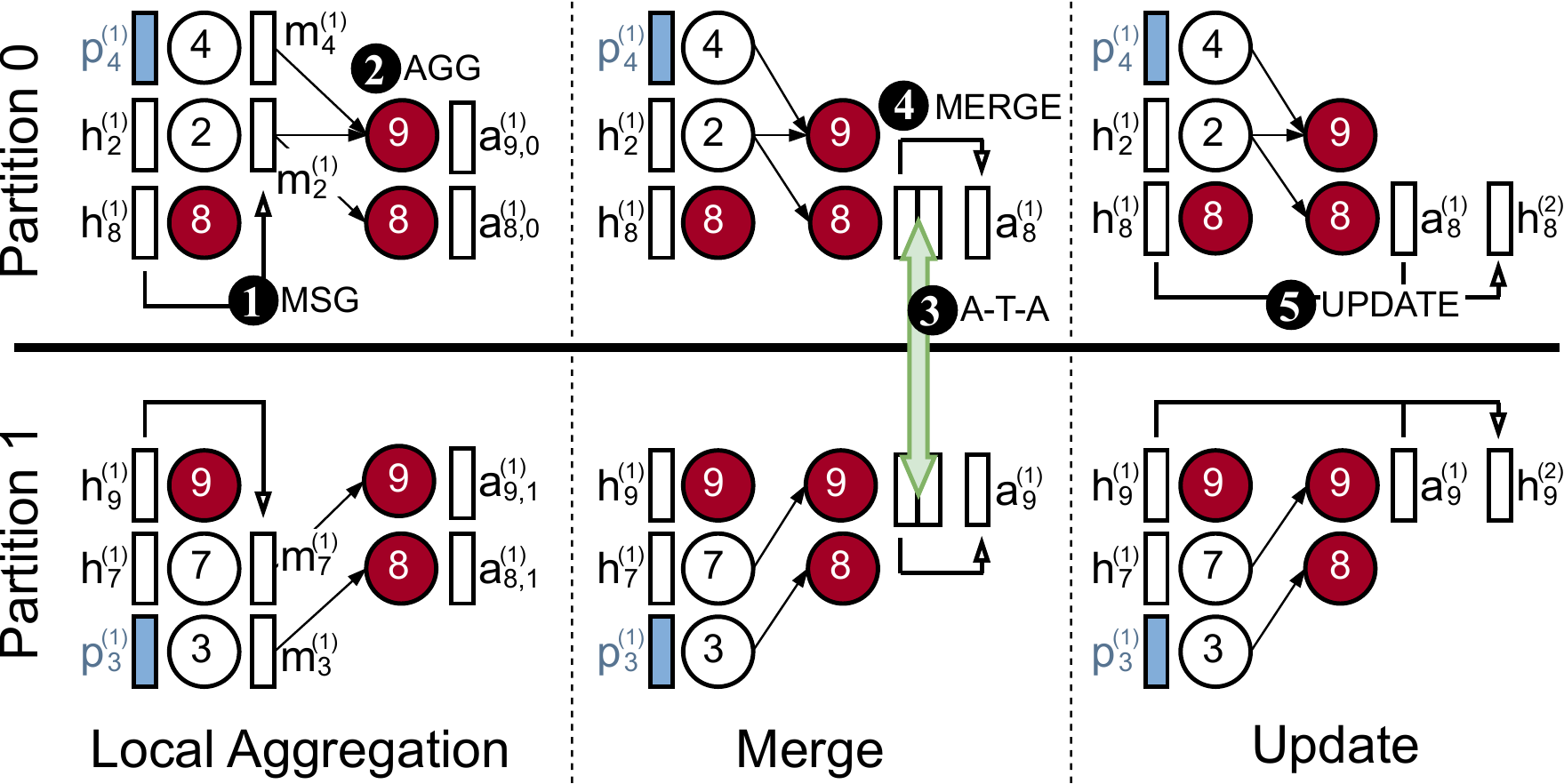}
  \caption{Example of distributed layer execution of CGP. At each partition, \numcircledtikz{1} the message function is applied to the input embeddings, and \numcircledtikz{2} the messages are aggregated by the local aggregation functions.
\numcircledtikz{3} The local aggregations of each node are shuffled through all-to-all communications.
Then, each partition proceeds to apply \numcircledtikz{4} the merge function and \numcircledtikz{5} the update function.
}
  \label{fig:cgp_2}
\end{figure}

Here, \smash{$P$} is the number of partitions, \smash{$\mathit{N}_p(v)$} returns the neighbors of a node \smash{$v$} in a partition \smash{$p$}, \smash{$a^{(l)}_{v, p}$} represents a local aggregation for \smash{$v$} in \smash{$p$}, \smash{$\hat{\bigoplus}$} is the local aggregation function, and \smash{$\biguplus$} is the merge function. In~\autoref{fig:cgp_2}, we depict the execution of the last layer in our running example.

\begin{algorithm}[t]
\scriptsize
\caption{\sys Distributed Layer Execution with Attention Mechanism}
\label{algo:attn}
\begin{algorithmic}[1]
\Statex \textbf{Params:} $M$: weights matrix, $\vec{a}$: learnable attention weights, $n\_parts$: number of partitions
\vspace{4pt}
\State \textbf{def }{Dist\_Layer\_Exec}{$(dst\_nodes, edges, embs)$:}
    \State \quad \textbf{for } $dst\_node \in dst\_nodes$ \textbf{ do} \Comment{Compute local aggregations}
        \State \quad \quad $h\_dst \gets M(embs[dst\_node])$
        \State \quad \quad $local\_srcs \gets \{ u | (u, dst\_node) \in edges \}$  
        \State \quad \quad $h\_local\_srcs \gets M(embs[local\_srcs])$
        \State \quad \quad $logits \gets \vec{a} (h\_dst || h\_local\_srcs)$
        \State \quad \quad $logit\_max \gets max(logits)$
        \State \quad \quad $exp\_sum \gets \sum_{l\in logits}{exp(l - logit\_max)}$
        \State \quad \quad $aggr \gets \sum_{u \in local\_srcs} \frac{exp(logits[u] - logit\_max)}{exp\_sum} h\_local\_srcs[u]$
    \State \quad \texttt{ALL-TO-ALL for local $aggr$, $exp\_sum$, and $logit\_max$ across partitions into $aggrs$, $exp\_sums$, and $logit\_maxs$}
    \State \quad \textbf{for } $dst\_node \in dst\_nodes$ \textbf{ do} \Comment{Merge local aggregations}
    \State \quad \quad $g\_logit\_max \gets \max(logit\_maxs)$
    \State \quad \quad $rescales \gets exp(logit\_maxs - g\_logit\_max)$
    \State \quad \quad $g\_exp\_sum \gets \sum_{p \in n\_parts} exp\_sums[p] * rescales[p]$
    \State \quad \quad $outputs[dst\_node]\gets \sum_{p \in n\_parts} \frac{aggrs[p] * rescales[p]}{g\_exp\_sum}$
    \State \quad \Return $outputs$
\end{algorithmic}
\end{algorithm}

\subsection{Custom Merge Functions}
\label{subsec:cgp_execution}

Applying CGP is straightforward for GNN models with commutative and associative aggregations, such as sum or max, but not so for more complex generalized arithmetic aggregations~\cite{pna2020,deepergcn2020} or softmax-based aggregation with learned weights~\cite{GAT2018,gat2_21}.
We explain how to handle them below.

\noindent
{\bf Generalized Arithmetic Aggregation.}
Beyond simple average functions, some GNN models leverage power mean aggregation~\cite{deepergcn2020} or normalized moments aggregation~\cite{pna2020}.
For power mean aggregation (\ie \smash{$(\frac{1}{|\mathit{N(v)}|}\sum_{u \in \mathit{N(v)}} m_u^p)^{\frac{1}{p}}$}), \sys computes the local aggregations by summing local messages after applying the \texttt{pow()} function with $p$.
To merge them, \sys adds the aggregations and then applies the \texttt{pow()} function with \smash{$1/p$}.
Normalized moments aggregation (\ie \smash{$(\frac{1}{|\mathit{N(v)}|}\sum_{u \in \mathit{N(v)}} (m_u - \bar{m})^n)^{\frac{1}{n}}$}) requires the mean of the messages ($\bar{m}$).
\sys first computes the mean values for each destination node (\eg nodes 8 and 9 in \autoref{fig:cgp_2}) then broadcasts the values with an all-gather operation.
Then, \sys computes the global aggregations with the same procedure used in the power mean aggregation.

\noindent
{\bf Softmax-based Aggregation.}
GNN models can utilize the attention mechanism~\cite{GAT2018,gat2_21}, which learns attention weights using the softmax function~\cite{softmax_1990} to identify important nodes in the neighborhood.
To employ the aggregation, a learned attention weight needs to be computed for each edge, which requires the embeddings of destination nodes~\cite{GAT2018}. For instance, in \autoref{fig:cgp_2}, the local aggregation in machine 0 for node 9 needs \smash{$h_9^{(1)}$}, which resides in machine 1. Consequently, as described in~\autoref{algo:attn}, \sys optionally employs an all-gather operation for destination embeddings (\ie \smash{$h_8^{(1)}$} and \smash{$h_9^{(1)}$}) to enable local aggregation.
Next, the attention weights are normalized using the softmax function.
For merging these softmax-based local aggregations while maintaining numerical stability, \sys additionally generates the exponential sum of logits and the maximum logits in each partition and adds them to the local aggregations, which are then utilized in the merging step. Recently, a similar technique has been applied to Transformer models~\cite{flash_attention_22}. We show that the accuracy impact of the aggregation is minimal in~\autoref{sec:overall_perf}.

We note there are some stateful aggregations~\cite{GraphSage2017,JKN2018} to which CGP cannot be directly applied;
e.g., those leveraging recurrent neural networks (RNNs)~\cite{lstm1997} to aggregate messages with recurrent hidden states. The strict dependency on aggregation prevents CGP from computing local aggregations in parallel. 
Since the RNN model must be sequentially applied to each neighbor in the resulting central aggregation, serving latency can become excessive. For instance, in our measurements of serving latency for a 3-Layer GraphSAGE~\cite{GraphSage2017} model with RNN-based and mean aggregations under identical settings in \autoref{tab:acc-lat-tradeoff}, we observe that the RNN-based aggregation requires substantially more time, taking 41$\times$ longer (28.9 s) compared to mean aggregation (702 ms). 
In future work, we will extend CGP and co-design GNN models to accommodate such aggregations, similar to learnable commutative aggregation~\cite{ong2022learnable}.

\section{Implementation}\label{sec:implementation}

We implement \sys in C++ and Python on top of the distributed Deep Graph Library (DGL)~\cite{DGL,DistDGL2020} and PyTorch~\cite{PyTorch}. Specifically, \sys reuses DGL's graph structures and message-passing APIs together with PyTorch's DNN operators and communication back-ends. Computation graph builders communicate via PyTorch GLOO~\cite{gloo}, while the master node coordinates with workers through PyTorch RPC~\cite{pytorch_rpc}. During execution, \sys customizes DGL's message-passing primitives to replace global aggregation with local aggregation, all-to-all communication, and merging. It uses DGL for local aggregation, PyTorch NCCL~\cite{NCCL} for cross-GPU communication, and PyTorch's DNN operations for the merging.

We adopt DGL because it is a widely adopted GNN framework in both academia and industry \cite{awsfraudgnn}. \sys runs unmodified GNN models written in DGL, as its executor is fully compatible with DGL and PyTorch APIs. Although we leverage these libraries for better usability, \sys's key techniques are library-agnostic and can be ported to other frameworks such as PyG~\cite{PyG2019}.

\begin{table}[t]
\centering
\resizebox{1.0\columnwidth}{!}{%
    \begin{tabular}{c|c|c|c|c|c}
        \toprule
        \textbf{Dataset} & \textbf{Nodes} & \textbf{Edges} & \textbf{Avg. Deg.} & \textbf{Features} & \textbf{Hiddens} \\
        \midrule
        Reddit~\cite{GraphSage2017} & 232~K & 115~M & 492 & 602 & 128 \\
        Yelp~\cite{graphsaint_20} & 717~K & 14.0~M & 20 & 300 & 512 \\
        Amazon~\cite{graphsaint_20} & 1.6~M & 264~M & 168 & 200 & 512 \\
        Products~\cite{OGB2020}  & 2.4~M & 124~M & 52 & 100 & 128 \\
        Papers~\cite{OGB2020} & 111~M & 1.6~B & 14 & 128 & 512 \\
        FB10B~\cite{FacebookGraph} & 30~M & 10~B & 333 & 1024 & 128 \\
        \bottomrule
    \end{tabular}
}
\caption{
Graph datasets used in the evaluation. The first four columns represent the number of nodes, number of directed edges, average degrees, and feature dimensions. The last column denotes the hidden dimensions used for GNN models.
}
\label{tab:exp-dataset}
\end{table}

\section{Evaluation}\label{sec:eval}
We evaluate the performance of \sys using popular benchmark graph datasets and representative GNN models.
We compare DGL-based~\cite{DGL,DistDGL2020} baseline serving systems with our techniques, SRPE (\autoref{sec:srpe}), CGP (\autoref{sec:cgp}), and their combination. Our evaluation addresses the following key questions: 
\begin{packeditemize}
    \item How much does \sys improve latency and accuracy compared to the baseline systems (\autoref{sec:overall_perf})?
    \item To what extent do SRPE and CGP contribute to the reduction in \sys's latency (\autoref{subsec:contribution_of_comps})? 
    \item What are the recomputation and memory overheads of using PEs (\autoref{subsec:overhead_of_pe})?
    \item How well does \sys scale with additional GPUs and machines (\autoref{subsec:scalability})?
    \item How do various system optimizations and model configurations affect \sys's latency (\autoref{subsec:sensitivity})?
\end{packeditemize}

\subsection{Experimental Setup}
\label{subsec:exp_setup}
\noindent
{\bf Testbed.}
We conducted our experiments on a GPU cluster with 4 servers, each equipped with two 32-core AMD 7542 CPUs, 512 GB of main memory, and two NVIDIA V100S GPUs with 32 GB of memory.
The servers are connected with 25~Gbps Ethernet links via a switch.
All servers run 64-bit Ubuntu 22.04, DGL v1.0.2, and PyTorch v1.13.0.

\noindent
{\bf GNN Models.}
We evaluate \sys on three representative GNN models, Graph Convolutional Networks (GCN)~\cite{GCN2017}, GraphSAGE (SAGE)~\cite{GraphSage2017}, and Graph Attention Networks (GAT)~\cite{GAT2018}. We use two layers for GCN and three layers for SAGE and GAT models to avoid the over-smoothing problem~\cite{gcn2_20,deepgcns_19,depper_insight_18} with deeper GNN models.

\noindent
{\bf Datasets.} We evaluate \sys on six graph datasets detailed in~\autoref{tab:exp-dataset}. Products~\cite{OGB2020} and Papers~\cite{OGB2020} are widely used for assessing GNN performance~\cite{P3,bgl_23,gnnautoscale_21,legion_23}. Reddit~\cite{GraphSage2017}, Yelp~\cite{graphsaint_20}, and Amazon~\cite{graphsaint_20} represent real-world web services where GNNs power applications like recommendations. For testing \sys on a larger scale, we include FB10B, a synthetic dataset with 10 billion edges modeled after Facebook's social network~\cite{FacebookGraph} with randomly generated 1,024-dimensional features, reflecting the high dimensionality in real-world scenarios~\cite{ogb-lsc,PinSage2018,AGL2020}.

\begin{table}[t]
\resizebox{1.0\columnwidth}{!}{%
\centering
\begin{tabular}{c|cc|cc|cc}
\toprule
\multirow{2}{*}{\textbf{Dataset}} 
& \multicolumn{2}{c|}{\textbf{GCN}} 
& \multicolumn{2}{c|}{\textbf{SAGE}} 
& \multicolumn{2}{c}{\textbf{GAT}} \\
& \textbf{Acc.~(\%)} & \textbf{$\gamma$~(\%)} 
& \textbf{Acc.~(\%)} & \textbf{$\gamma$~(\%)} 
& \textbf{Acc.~(\%)} & \textbf{$\gamma$~(\%)} \\
\midrule
\textbf{Reddit} & 92.5~(-0.1) & 0 & 96.3~(0.0) & 0 & 95.5~(0.0) & 0 \\
\textbf{Yelp} & 42.4~(-4.5) & 20 & 63.7~(0.0) & 0 & 56.9~(-6.3) & 7 \\
\textbf{Amazon} & 41.8~(-3.9) & 3 & 79.4~(0.0) & 0 & 63.4~(-3.0) & 1 \\
\textbf{Products} & 75.6~(+0.1) & 0 & 77.8~(+0.1) & 0 & 73.3~(0.0) & 0 \\
\textbf{Papers} & 42.0~(+1.4) & 0 & 50.0~(+0.3) & 0 & 49.3~(+0.4) & 0 \\
\bottomrule
\end{tabular}}
\caption{
  Effectiveness of \sys's recomputation policy (\autoref{subsubsec:policy}). `Acc.' column represents the accuracy with full computation graphs and the percentage of accuracy drops with PEs (without recomputation). `$\gamma$' column shows the recomputation budget to achieve less than 1\% points of accuracy drop. We randomly remove 25\% of test nodes to create serving requests, each containing 1,024 query nodes, and compute PEs using the remaining nodes. Accuracy drops and the recomputation budgets are then aggregated across these requests. We train the models using grid search on learning rates (0.01, 0.001, 0.0001) with the Adam optimizer~\cite{kingma2014adam} and dropout probabilities (0.1, 0.5), running fixed numbers of epochs based on convergence. We use neighborhood sampling in training with (25, 10) and (15, 10, 5) fanouts.
}
\label{tab:srpe_table}
\end{table}

\begin{figure*}[t]
\centering
  \includegraphics[width=\textwidth]{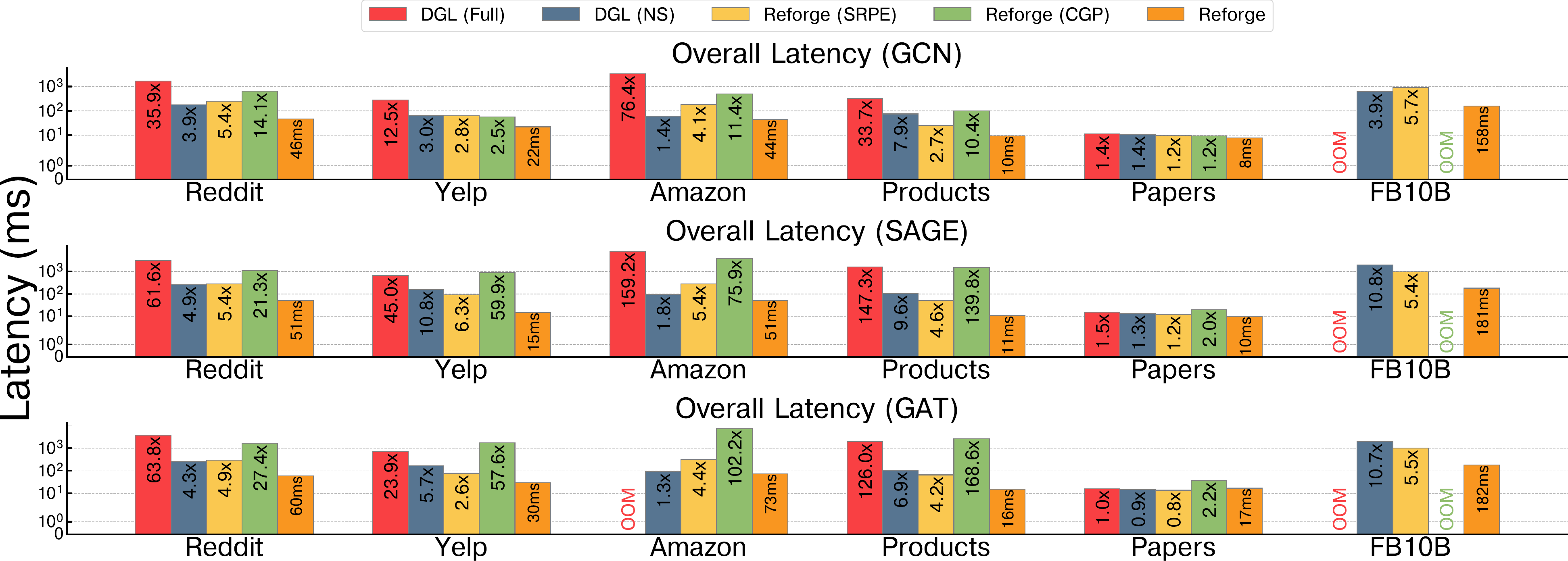}
  \caption{End-to-end serving latency of \sys and baseline systems (in log scale) across three models and six datasets. The numbers within each bar represent the latency values for \sys or the relative speedup of \sys compared to each system. `OOM' indicates that a system failed to execute the workload due to a CUDA Out-of-Memory error.}
  \label{fig:overall_latency}
\end{figure*}

\noindent
{\bf Workloads.}
We synthesize realistic serving workloads from the datasets since no public large-scale GNN serving workload is available.
For each dataset, we remove 25\% of random test nodes and the edges connected to the nodes.
We make a serving request by randomly selecting a specific number of query nodes from the removed nodes and the edges from the query nodes to the nodes in the remaining dataset.
Our evaluation uses batch sizes of 64, 128, 256, 512, 1,024, and 2,048.
For each batch size, we generate 500 serving requests (reused for all evaluations).
We execute one request at a time to measure the serving latency without any resource contention among requests. 
The reported results are averaged over the 500 serving requests.

\noindent
{\bf Baselines.}
For a fair comparison with \sys, which is based on Distributed DGL (\autoref{sec:implementation}), we implement the following two baseline GNN service systems using DGL~\cite{DistDGL2020}.\footnote{Since DGL is primarily designed for training, we adapt distributed DGL~\cite{DistDGL2020} by removing the backward pass and using only the forward pass in both baselines.}
\begin{packeditemize}
    \item {\textsf{\textbf{DGL~(Full)}}} constructs and executes full computation graphs that consist of the entire $k$-hop neighborhood of query nodes. This approach is typically preferred for relatively small graphs~\cite{lambdagrapher,fograph}.
    \item {\textsf{\textbf{DGL~(NS)}}} employs neighborhood sampling to reduce latencies at the expense of accuracy. This approach is essential today for supporting large-scale graphs in existing GNN serving systems~\cite{PinSage2018,quiver2023,lin2022platogl}. We use widely adopted sampling fanouts~\cite{Salient2022,GraphSage2017,quiver2023,helios} of (25, 10)\footnote{The (25, 10) fanout means sampling at most 10 neighbors at the first hop and sampling at most 25 neighbors for each direct neighbor.} and (15, 10, 5) for models with 2 and 3 layers unless otherwise specified.
\end{packeditemize}

\noindent
{\bf Default Configurations.} Unless otherwise specified, experiments use a batch size of 1,024 and run on 4 machines, each with 1 GPU. \sys employs the recomputation policy (\autoref{subsubsec:policy}), with budgets ($\gamma$) set for less than 1\% accuracy drop (\autoref{tab:srpe_table}). For the FB10B dataset, containing synthetic feature vectors, no recomputation is performed. Random hash partitioning is applied by default for better load balancing, as locality-aware strategies like Metis~\cite{METIS} do not improve \sys or baseline performance (\autoref{subsec:sensitivity}).

\subsection{Overall Performance} \label{sec:overall_perf}
We evaluate the end-to-end serving latency and accuracy of \sys in comparison to \dglf and \dglns. In \autoref{fig:overall_latency}, we report the latencies of \sys and the baseline systems, along with relative speedups, across six datasets listed in \autoref{tab:exp-dataset} and GCN, SAGE, and GAT models. To highlight the contributions of SRPE and CGP, we also include latencies for \syssrpe and \syscgp, representing \sys with only SRPE and CGP, respectively.

\noindent
{\bf Compared to {\textsf{\textbf{DGL~(Full)}}}.}
\sys significantly reduces latency while minimizing accuracy drop.
For instance, as shown in~\autoref{fig:overall_latency}, \textit{\sys achieves 159$\times$ lower latency than \dglf for the SAGE model with the Amazon dataset}.
\sys handles large graphs with higher degrees efficiently, leveraging SRPE's precomputation and CGP's local aggregation.
For the FB10B dataset, \sys successfully runs GNNs on the largest dataset by significantly reducing computation graph sizes, whereas \dglf fails due to CUDA out-of-memory (OOM) errors.
The Papers dataset shows little benefit from \sys since its serving nodes have low degrees (2.4 on average), and \dglf shows low latency of 15~ms.

\noindent
{\bf Compared to {\textsf{\textbf{DGL~(NS)}}}.}
We first study how neighborhood sampling affects accuracy. For GCN and GAT models on Amazon and Yelp datasets, accuracy loss from sampling compared to \sys is significant (1.7\% to 5.0\% in \autoref{tab:sampling_acc_loss}), while \sys achieves substantial latency speedups (1.3$\times$ to 5.7$\times$ in \autoref{fig:overall_latency}) with less than 1\% point accuracy loss. SAGE models are more resilient to sampling, typically showing negligible accuracy losses, as reported in prior work~\cite{Salient2022}. Overall, \sys demonstrates minimal accuracy drop while significantly reducing latency across all cases.

Additionally, sampling allows \dglns to handle the largest dataset, FB10B, without CUDA OOM. However, even with sampling, the total neighbors can be large. For instance, with a (15, 10, 5) sampling configuration, one query node can have up to 750 3-hop neighbors, leading to high communication overheads for fetching associated feature vectors, resulting in a latency of 2.0 seconds for FB10B.
In contrast, \textit{\sys effectively reduces computation graph sizes through SRPE and minimizes communication overheads with CGP, outperforming \dglns by up to $10.8\times$.}

\begin{table}[t]
\centering
\resizebox{0.95\linewidth}{!}{%
\centering
\begin{tabular}{c|ccc|ccc}
\toprule
\multirow{2}{*}{\begin{tabular}[c]{@{}c@{}}\textbf{Systems}\end{tabular}}
 & \multicolumn{3}{c}{\textbf{Yelp}} &
 \multicolumn{3}{c}{\textbf{Amazon}} \\
  & \textbf{GCN} & \textbf{SAGE} & \textbf{GAT}
                    & \textbf{GCN} & \textbf{SAGE} & \textbf{GAT} \\ 
\midrule
\dglns & 36.2\% & 63.7\% & 50.9\% & 36.1\% & 79.4\% & 60.8\% \\
\sys & 41.5\% & 63.7\% & 55.9\% & 40.9\% & 79.4\% & 62.5\% \\
\bottomrule
\end{tabular}}
    \caption{
    Test accuracies of \sys and \dglns using three GNN models (GCN, SAGE, GAT) on the Yelp and Amazon datasets. \sys employs SRPE with the recomputation budgets from \autoref{tab:srpe_table}, maintaining less than 1~\% point of accuracy loss. Corresponding latency results are provided in \autoref{fig:overall_latency}.
    }
    \label{tab:sampling_acc_loss}
\end{table}

\subsection{Contributions of \sys's Techniques}
\label{subsec:contribution_of_comps}
To better understand the performance benefits of \sys, we analyze the respective contributions of SRPE and CGP.

\begin{figure}[t]
  \centering
  \includegraphics[width=1.0\linewidth]{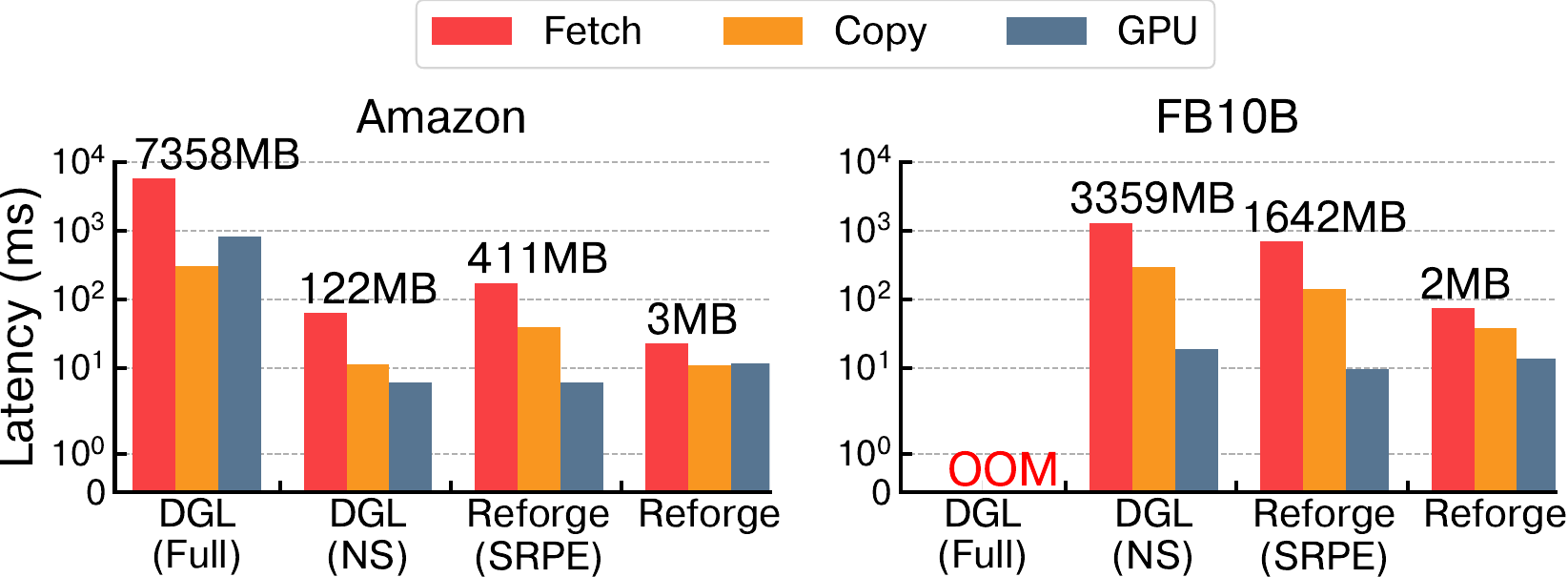}
  \caption{
    Latency breakdown of SAGE model with Amazon and FB10B datasets in \autoref{fig:overall_latency}, which consists of three components:
    `Fetch' builds computation graphs by fetching necessary data (\ie remote feature vectors, edges, and PEs), `Copy' transfers the data into GPU device memory, and `GPU' runs GNN computations in GPUs (including collective communications for \sys's CGP). The size of data in fetching and collective communications is noted on top of each bar.}
  \label{fig:overall-breakdown}
\end{figure}

\noindent
{\bf Contribution of SRPE.}
We present the latency breakdown and required communication size of \dglf, \dglns, \syssrpe, and \sys in~\autoref{fig:overall-breakdown}.
\dglf suffers from fetching feature vectors and edges for the entire $k$-hop neighborhood during computation graph creation (\texttt{Fetch}), taking several seconds to serve a single request or causing CUDA OOM.
In contrast, SRPE reduces computation graph sizes and communication overhead. For example, in the Amazon dataset, \textit{SRPE reduces the communication volume by 18$\times$, from 7.4GB to 411MB, and latency by 29$\times$, from 8.2s to 278ms}.

However, compared to \dglns, communication size in \syssrpe can still be significant, as shown with the Amazon dataset in~\autoref{fig:overall-breakdown}. This occurs when query nodes have large numbers of direct neighbors (\eg 168 average degrees in~\autoref{tab:exp-dataset}) and PEs are larger than feature vectors (\eg 512 hidden vs. 200 feature dimensions for Amazon in~\autoref{tab:exp-dataset}).

\noindent
{\bf Contribution of CGP.}
CGP's local aggregation effectively reduces the communication overhead of SRPE, as it only requires collective communications for the query nodes and a small number of recomputation target nodes (\eg nodes 2, 7, 8, and 9 in~\autoref{fig:cgp_1b}).
As shown in~\autoref{fig:overall-breakdown} (denoted \sys), \textit{CGP's local aggregations minimize communication to a few MBs of necessary collective communications, reducing latency of \syssrpe by 5.5$\times$ (from 278~ms to 51~ms) and 5.4$\times$ (from 978~ms to 181~ms)} for the Amazon and FB10B datasets.

We observe that CGP alone can provide limited benefit. As \autoref{fig:overall_latency} shows (denoted \syscgp), while CGP reduces latency for 2-hop computation graphs (\ie GCN) relative to \dglf, its performance for 3-hop graphs (\ie SAGE, GAT) is poorer since deeper GNN models' neighborhoods expand to encompass most of a dataset, limiting local aggregation effectiveness.

\begin{figure}[t]
  \centering
  \includegraphics[width=1.0\linewidth]{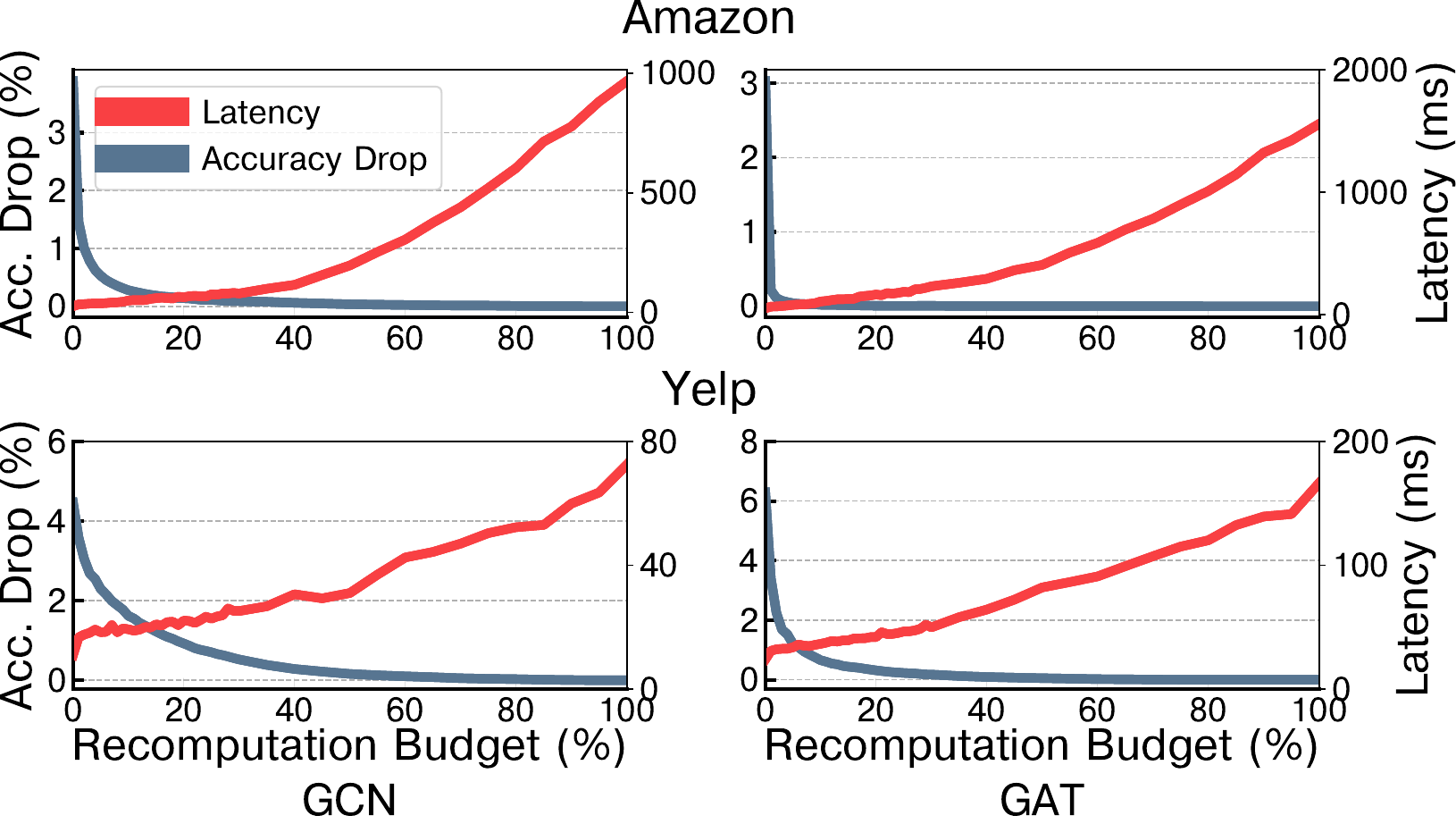}
  \caption{Latency and accuracy trade-off of \sys varying recomputation budget with GCN and GAT models and Amazon and Yelp datasets.}
  \label{fig:recomputation}
\end{figure}

\subsection{Recomputation and Memory Overhead of PEs}
\label{subsec:overhead_of_pe}
While SRPE provides significant latency benefits, it also introduces additional costs: recomputation (to mitigate accuracy losses) and memory (to store PEs on the host). We evaluate the impact of recomputation on latency and accuracy, and discuss the memory overhead of PEs.

\noindent
{\bf Recomputation Overhead.}
We assess the trade-offs between latency and accuracy based on the budget of \sys's recomputation policy  (\autoref{subsec:recom_policy}). \autoref{fig:recomputation} presents results for GCN and GAT models using the Yelp and Amazon datasets, which experience the largest accuracy drops without PE recomputation.
\textit{The results demonstrate that \sys's policy effectively minimizes recomputation costs; for instance, a recomputation budget of 7\% reduces accuracy losses from 6.3\% points to just 1.0\% points, while increasing latency by only 11 ms for the GAT model on the Yelp dataset.}

\noindent
{\bf Memory Overhead.}
The memory required to store PEs is proportional to the number of layers, hidden dimensions, and data type size: $(L - 1) * H * D$ bytes. Although the memory footprint depends on the GNN model configuration and graph dataset size, \textit{PEs are typically smaller than the graph dataset size, as models generally have only 2–3 layers and relatively small hidden dimensions~\cite{P3}. For instance, the memory footprint of PEs for the 3-layer SAGE model is 31.5 GB, representing just 10.9\% of the FB10B dataset size.}
Among the evaluated workloads, the Papers dataset results in the largest PE memory footprint. For a 3-layer GNN with 512-dimensional hidden states, storing PEs requires approximately 455 GB. \sys stores PEs in host CPU memory and distributes them across machines using CGP; in our 4-machine evaluation cluster with 2 TB of aggregate host RAM, this corresponds to 22\% of the available memory. This enables \sys to eliminate recursive neighborhood fetching during serving, yielding order-of-magnitude latency reductions (\autoref{fig:overall_latency}).

\begin{figure}[t]
  \centering
  \includegraphics[width=1.0\linewidth]{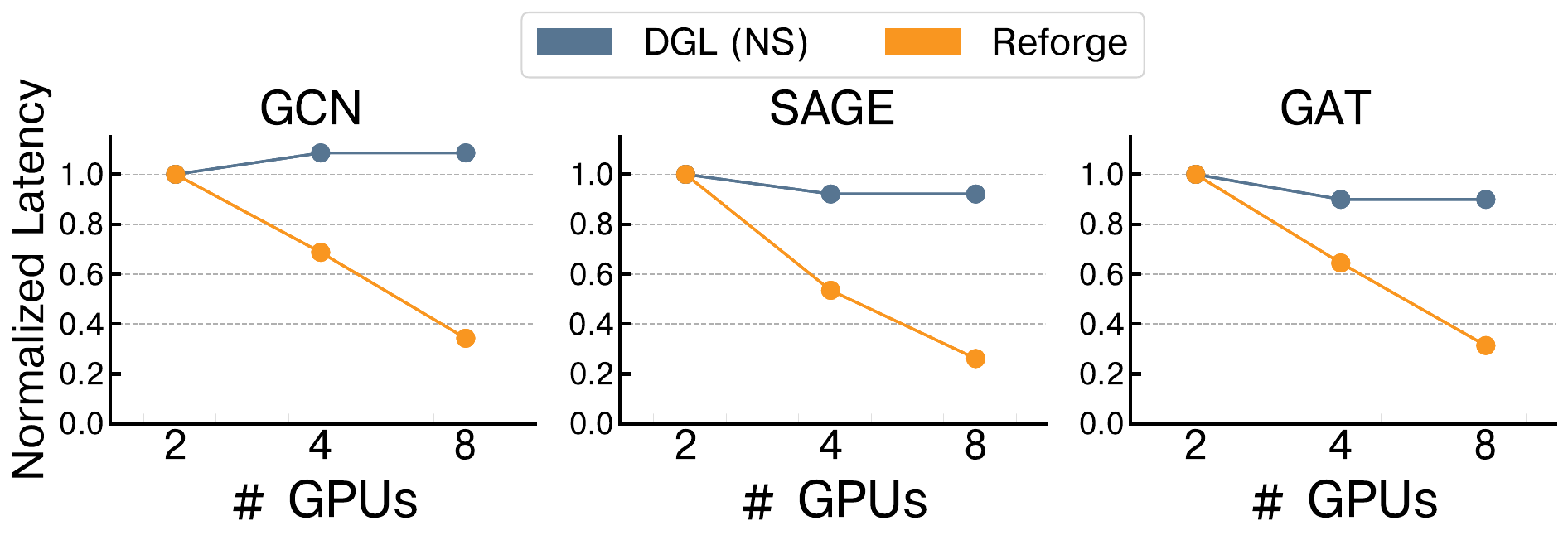}
  \caption{Normalized Latency of \sys and \dglns varying number of GPUs with FB10B dataset.}
  \label{fig:scalability}
\end{figure}

\begin{figure}[t]
  \centering
        \includegraphics[width=\linewidth]{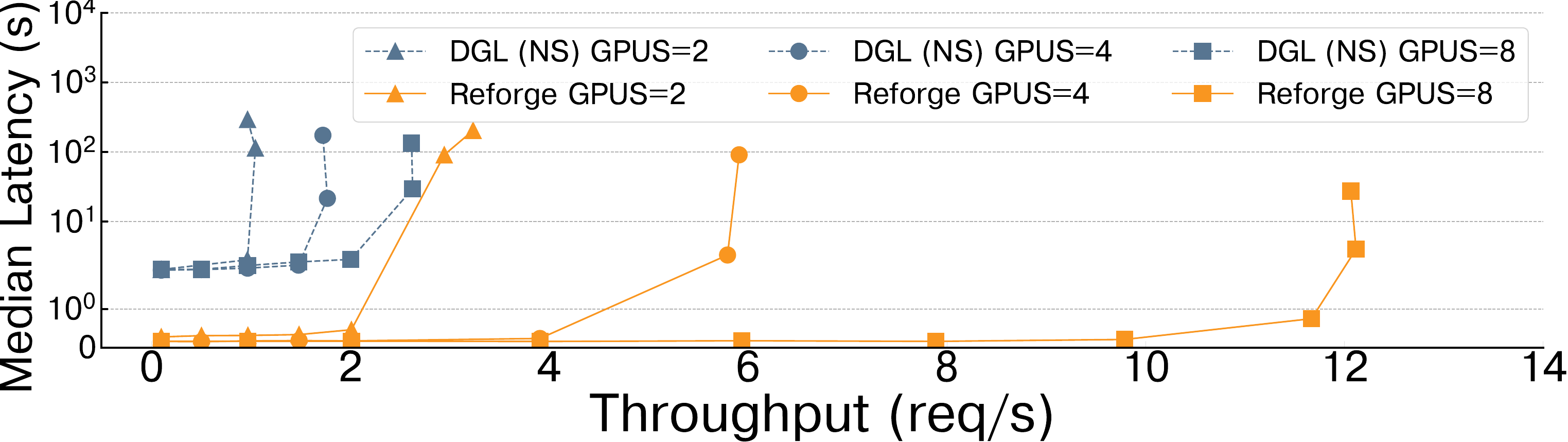}
        \caption{Latency-throughput results of \sys and \dglns varying request rates with SAGE model and FB10B dataset.}
        \label{fig:throughput}
\end{figure}

\subsection{Scalability Analysis}
\label{subsec:scalability}
In this section, we examine the impact of varying the number of machines and GPUs on the latency and throughput of \sys. We evaluate the largest FB10B dataset, and compare \sys with \dglns since \dglf cannot run the workload due to CUDA OOM.

\noindent
{\bf Latency.}
In \autoref{fig:scalability}, we vary the number of GPUs: 2 (on 2 machines), 4 (on 4 machines), and 8 (with 4 machines each using 2 GPUs).
While \dglns does not benefit from additional resources due to its centralized execution, \sys demonstrates strong scaling by distributing the host-to-GPU memory transfer across multiple GPUs, with each GPU handling only local feature vectors and PEs.
\textit{As a result, for instance, the latency of \sys with the GAT model decreases by 67\% (from 282ms to 88ms), whereas \dglns only shows a 9\% (from 2.2s to 2.0s) reduction in latency.}

\noindent
{\bf Throughput.}
We analyze the throughput of \sys and \dglns by modeling request arrivals with a Poisson distribution and feeding workloads into both systems in~\autoref{fig:throughput}. While \dglns enables concurrent request handling by individual GPUs, significant network contention limits its scalability, achieving only a 2.6$\times$ increase in maximum throughput from 2 to 8 GPUs. In contrast, \textit{\sys leverages CGP to reduce communication overhead and avoid contention, resulting in a 3.8$\times$ increase in throughput. Furthermore, with 8 GPUs, \sys outperforms \dglns by 4.7$\times$ while maintaining significantly lower latency.}

\subsection{Impact of Optimizations and Configurations}
\label{subsec:sensitivity}
We evaluate the impact of system optimizations and model configurations on \sys's latency. To minimize communication and host-to-device transfer overheads, GNN training systems often utilize GPU feature caching~\cite{PaGraph2020, gnnlab_22, bgl_23, legion_23} and locality-aware partitioning~\cite{PaGraph2020, bgl_23, DistDGL2020, legion_23, bytegnn_22, distgnn_21}. We analyze these techniques' effects on \sys's latency and explore various model configurations.

\begin{figure}[t]
  \centering
        \includegraphics[width=\linewidth]{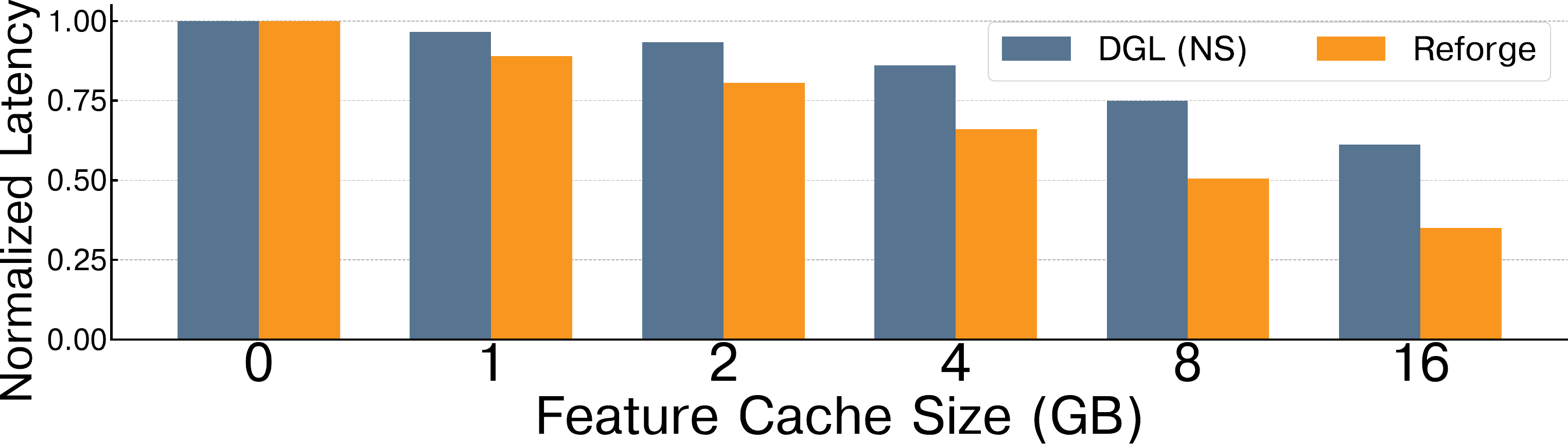}
        \caption{Normalized latency of \sys and \dglns varying feature cache size with SAGE model and FB10B dataset.}
        \label{fig:feature_cache}
\end{figure}

\begin{table}[t]
\resizebox{\columnwidth}{!}{%
\centering
\begin{tabular}{c|cc|cc}
\toprule
\textbf{Systems} &
\textbf{Yelp + R.H.} &
\textbf{Yelp + Metis} &
\textbf{Amazon + R.H.} &
\textbf{Amazon + Metis} \\ 
\midrule
\textsf{\textbf{DGL~(Full)}} &
652.6 $\pm$ 30.8 &
774.2 $\pm$ 37.4 &
8181.3 $\pm$ 215.4 &
10602.8 $\pm$ 1087.6 \\
\textsf{\textbf{DGL~(NS)}} &
159.1 $\pm$ 3.2 &
167.0 $\pm$ 8.4 &
93.0 $\pm$ 5.4 &
100.1 $\pm$ 11.4 \\
\textbf{\sys} &
15.2 $\pm$ 0.5 &
15.3 $\pm$ 0.9 &
49.1 $\pm$ 3.3 &
86.7 $\pm$ 11.0 \\
\bottomrule
\end{tabular}}
\caption{Serving latency and standard deviation (in milliseconds) for two graph partitioning strategies (Random-Hash vs. Metis) with Yelp and Amazon datasets.}
\label{table:graph_partitioning}
\end{table}

\noindent
{\bf Impact of Feature Cache.}
We evaluate how the feature caches in GPUs impact the serving latency of \sys and \dglns. 
Following prior work~\cite{PaGraph2020}, we sort nodes by their out-degrees, caching the highest-ranked nodes to maximize the chances of cache hits.
As shown in \autoref{fig:feature_cache}, \textit{both \sys and \dglns benefit from caching, but \sys sees a much larger reduction in latency. For example, with a 16GB cache, \sys's latency drops by 65\%, compared to 39\% for \dglns.} This is because \sys takes advantage of CGP's distributed execution, where each GPU caches frequently accessed nodes from its local subset of data. In contrast, \dglns uses a centralized execution approach, requiring every GPU to consider nodes from the entire dataset.

\noindent
{\bf Impact of Graph Partitioning.}
We study the impact of graph partitioning on serving latency of \sys and the baseline systems using two partitioning strategies: Metis~\cite{METIS}, a widely-used locality-aware partitioning method, and random-hash partitioning. As shown in \autoref{table:graph_partitioning}, \textit{we observe that all systems show reduced latency and variance with random hash partitioning compared to using Metis.} This suggests that locality-aware partitioning, while effective during the training phase for reducing communication costs, does not provide similar benefits for serving workloads, as query nodes do not fully leverage the locality established before they arrive.

\begin{figure}[t]
  \centering
  \includegraphics[width=\linewidth]{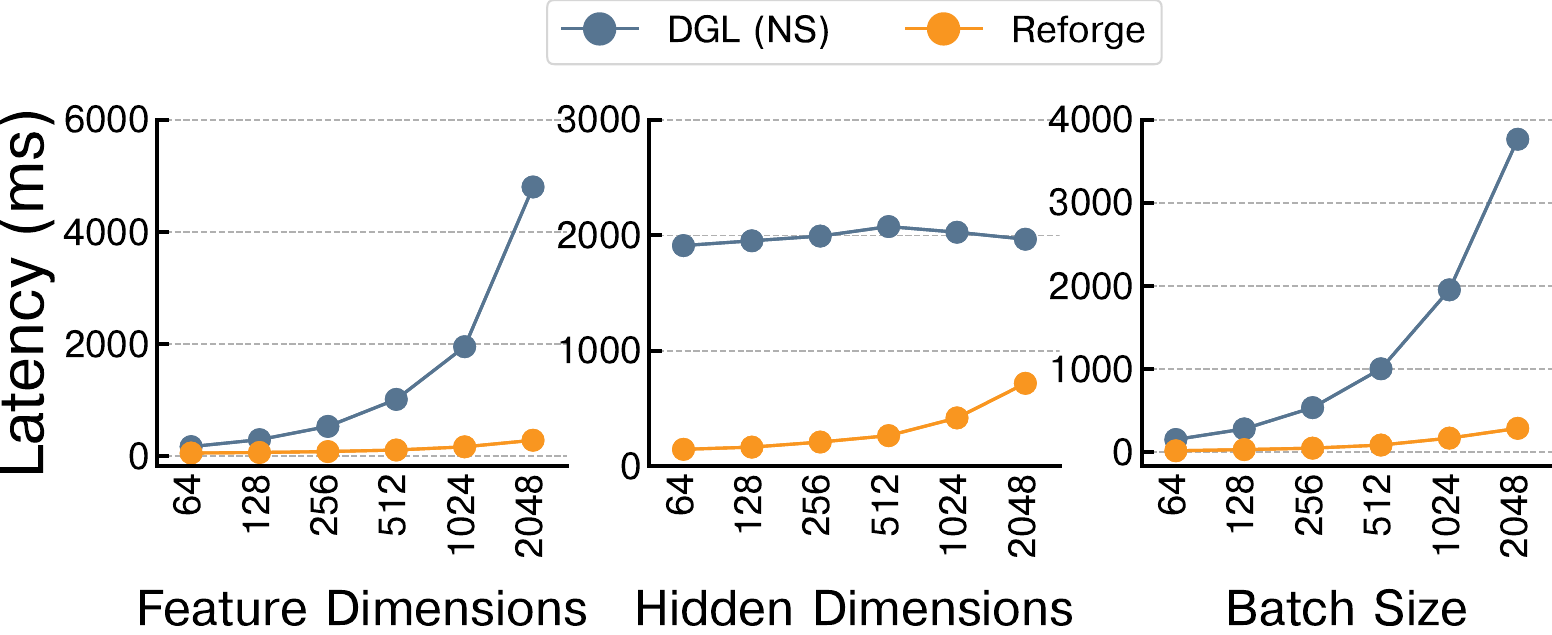}
  \caption{Latency of \sys and \dglns varying feature dimensions, hidden dimensions, and batch sizes with SAGE model and FB10B dataset.}
  \label{fig:params}
\end{figure}

\noindent
{\bf Impact of Model Configurations.}
Finally, we assess the impact of model configurations on the serving latency of \sys.
In~\autoref{fig:params}, we analyze the serving latency of \sys on the FB10B dataset using the SAGE model, varying feature dimensions, hidden dimensions, and batch sizes. The default configuration is (1024, 128, 1024).
We observe that \textit{\sys effectively handles larger configurations, outperforming \dglns by $16.9\times$ and $13.3\times$ for feature dimension and batch size of 2048, respectively.}
While larger hidden dimensions increase \sys's latency—since it communicates layer embeddings of hidden size (\autoref{subsec:cgp_creation})—\sys still achieves $2.7\times$ lower latency at 2048 hidden dimensions, benefiting from reduced communication costs enabled by SRPE and CGP.

\begin{figure}[t]
  \centering
  \includegraphics[width=\linewidth]{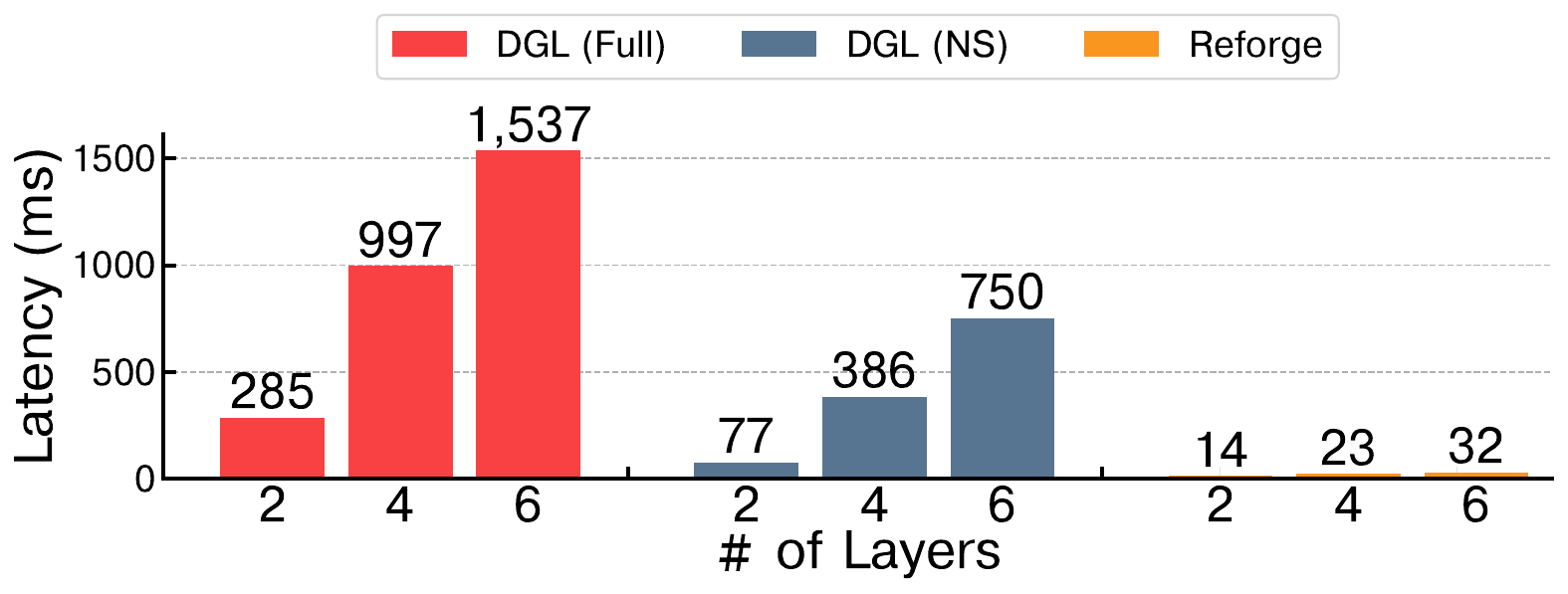}
  \caption{Latency of \sys and \dglns varying number of layers with GCNII model and Yelp dataset.}
  \label{fig:layer}
\end{figure}

We further evaluate the effect of increasing the number of GNN layers using the GCNII model~\cite{gcn2_20}, which is designed to mitigate the over-smoothing problem in deep GNNs~\cite{gcn2_20,deepgcns_19,depper_insight_18}.
As shown in~\autoref{fig:layer}, \textit{\sys exhibits linear latency scaling with respect to the number of layers, whereas the baselines suffer from exponential growth.} This allows \sys to outperform \dglf and \dglns by $20.3\times$ and $5.5\times$ at 2 layers, and by $48.0\times$ and $23.4\times$ at 6 layers, respectively. This is due to PEs allowing \sys to construct computation graphs only with direct neighbors, causing the size of the computation graph to grow only linearly with the number of layers.

\section{Related Work}\label{sec:related}

\noindent
{\bf GNN Serving Systems.}
Several systems~\cite{PinSage2018,quiver2023,lin2022platogl,helios} accelerate GNN serving for large graphs through sampling. However, as discussed in \autoref{subsubsec:acc_drop} and \autoref{sec:overall_perf}, sampling often results in significantly higher latency and/or lower accuracy. Other systems~\cite{lambdagrapher,fograph} focus on resource-efficient techniques for optimizing GNN serving in decentralized environments. While techniques such as adaptive batching~\cite{lambdagrapher} can be integrated with \sys, serving with the full computation graphs in these systems is prohibitively memory-intensive and slow (\autoref{sec:overall_perf}). In contrast, \sys's SRPE efficiently reduces computation graph size through selective recomputation, achieving higher accuracy than sampling-based approaches while requiring smaller graphs (\autoref{sec:overall_perf}). Additionally, while existing GPU-based GNN serving systems such as Quiver~\cite{quiver2023} and SALIENT~\cite{Salient2022} rely on centralized execution with feature caching limited to single-GPU memory, \sys employs CGP to enable distributed execution with pooled GPU memory across the cluster, which improves the effectiveness of feature caching at scale (\autoref{subsec:sensitivity}).

RIPPLE~\cite{ripple_25} and InkStream~\cite{inkstream_23} incrementally update node embeddings from the deltas caused by query nodes to address the overhead of GNN serving. However, those systems primarily support linear aggregation and have limited support for attention-based models. In contrast, \sys systematically employs CGP with custom merge functions, supporting a wide range of GNN models.

\noindent
{\bf GNN Kernel Optimizations.}
Existing systems optimize GNN execution on GPUs using techniques like kernel fusion and pipelining to enhance performance~\cite{NeuGraph2019, gnn_advisor_21, mgg_23, seastar_21}.
For example, GNNAdvisor~\cite{gnn_advisor_21} reduces atomic operations and global memory access through runtime graph-aware warp and block-level organization.
While these optimizations could be integrated into \sys, they do not address communication overheads in serving.
Similarly, HAG~\cite{HAG2020} introduces hierarchical aggregation to eliminate redundancy, which is comparable but distinct from CGP's local aggregation, presenting an interesting future research direction.

\noindent
{\bf GNN Training Systems.}
To mitigate the overhead of large neighborhoods, many GNN training systems use approximations such as historical embeddings~\cite{gnnautoscale_21,sancus_22,freshgnn} or sampling~\cite{PyG2019,DGL,PaGraph2020,gnnlab_22,bgl_23,legion_23,P3,AGL2020,AliGraph2019,DistDGL2020,leapgnn}. However, both approaches often fall short of preserving full model accuracy. Historical embeddings do not account for dependencies introduced by query nodes, degrading accuracy (\autoref{tab:srpe_table}), while sampling approximates the full neighborhood and similarly causes accuracy loss (\autoref{tab:sampling_acc_loss}). Moreover, these systems rely on centralized execution, leading to high latency in GNN serving (\autoref{fig:overall_latency}). In contrast, \sys achieves minimal accuracy loss by SRPE and leverages distributed resources to reduce latency via CGP, significantly outperforming the approximation techniques in both accuracy and latency (\autoref{sec:overall_perf}).

On the other hand, full-graph training systems~\cite{NeuGraph2019, ROC2020, bnsgcn_22, dorylus_21} avoid computation graph construction by operating on the entire graph. However, as they are designed for static graphs and lack dynamic computation graph creation for new query nodes, their techniques are either inapplicable to GNN serving or complementary to \sys.

\section{Conclusion}\label{sec:conclusion}
This paper presents \sys, a GNN serving system achieving low latency and minimal accuracy loss for large graphs. \sys addresses neighborhood explosion with SRPE~(\autoref{sec:srpe}), which precomputes layer embeddings and selectively recomputes error-prone embeddings in the serving phase. \sys further reduces communication overhead through CGP~(\autoref{sec:cgp}), enabling distributed computation graph creation with custom merge functions. Our evaluation demonstrates that \sys reduces latency by orders of magnitude compared to baselines with minimal accuracy loss~(\autoref{sec:eval}).

\bibliographystyle{IEEEtran}
\bibliography{references}

\appendices

\section{Proof of \autoref{theorem:minvar}}
\label{appendix:proof}

In this section, we describe the formal proof of \autoref{theorem:minvar}. To prove the theorem, we assume each GNN layer independently learns embeddings, following prior works~\cite{AsGCN2018,graphsaint_20,FastGCN2018}. This assumption allows
GNN models to be statistically analyzed despite complex nonlinear activations between layers.
Then, as we described in \autoref{subsubsec:problem_formulation}, for each recomputation candidate $u \in \mathit{R}$, we have an unbiased estimator
\begin{gather*}
    \hat{f}_u^{(l)} = \frac{1}{p_u}\hat{z}_u q_u^{(l)} + t_u^{(l)}, \\
    q_u^{(l)} = \sum_{v \in \mathit{N_Q}(u)} \frac{m_v^{(l)}}{|\mathit{N}(u)|}, \quad
    t_u^{(l)} = \sum_{v \in \mathit{N_T}(u)} \frac{m_v^{(l)}}{|\mathit{N}(u)|},
\end{gather*}
where $N_Q(u)$ and $N_T(u)$ represent neighbors of $u$ in query nodes and training nodes, respectively.

Let the embeddings be $d$-dimensional vectors.
We find the optimal recomputation probabilities ($p_u$), given recomputation budget $\gamma = \sum_{u \in \mathit{R}}p_u$, that minimize the following sum of variances.
\def\Var{{\textrm{Var}}\,}
\def\Cov{{\textrm{Cov}}\,}
\begin{equation}
    S = \sum_{i=1}^d\Var[ \sum_{u\in \mathit{R}} \sum_{l=1}^{k-1} (\frac{1}{p_u}\hat{z}_u q_u^{(l)} + t_u^{(l)})_i]
\end{equation}
By the independence of recomputation variables ($\hat{z}_u$), we have $\Cov[\hat{z}_u, \hat{z}_v] = 0$ if $u \neq v$. Also, $\Cov[\hat{z}_u, \hat{z}_u] = p_u(1-p_u)$. Thus,
\begin{equation}
\begin{aligned}
    S &= \sum_{i=1}^d \sum_{u \in \mathit{R}} (\sum_{l=1}^{k-1}(q_u^{(l)})_i)^2 \frac{1}{p_u^2}p_u(1-p_u) \\
      &= \sum_{u \in \mathit{R}} \sum_{i=1}^d (\sum_{l=1}^{k-1}(q_u^{(l)})_i)^2 (\frac{1}{p_u} - 1) \\
      &= \sum_{u \in \mathit{R}} || \sum_{l=1}^{k-1} q_u^{(l)} ||^2 (\frac{1}{p_u} - 1)
\end{aligned}
\end{equation}

Now, it is sufficient to minimize the first term
\begin{equation*}
    \sum_{u \in \mathit{R}} || \sum_{l=1}^{k-1} q_u^{(l)} ||^2 \frac{1}{p_u}.
\end{equation*}
By Cauchy-Schwarz inequality,
\begin{equation}
\begin{aligned}
\Big(\sum_{u \in \mathit{R}} \big(|| \sum_{l=1}^{k-1} q_u^{(l)} || \frac{1}{\sqrt{p_u}}\big)^2\Big)\Big(\sum_{u \in \mathit{R}} \sqrt{p_u}^2\Big) \\
\geq \Big(\sum_{u \in \mathit{R}}|| \sum_{l=1}^{k-1} q_u^{(l)} ||\Big)^2
\end{aligned}
\end{equation}
Since $\gamma = \sum_{u \in \mathit{R}} \sqrt{p_u}^2$ and the right-hand side are constants, the first term is minimized when the following equality condition holds:

\begin{equation}
\forall u \in \mathit{R}, || \sum_{l=1}^{k-1} q_u^{(l)} || \frac{1}{\sqrt{p_u}} \propto \sqrt{p_u}
\end{equation}

Therefore, we conclude $S$ is minimized if $p_u \propto || \sum_{l=1}^{k-1} q_u^{(l)} ||$.

\begin{table}[t]
\small
\centering
\resizebox{\columnwidth}{!}{%
    \begin{tabular}{c|c|c|c|c|c|c|c|c}
        \toprule
        \textbf{Dataset} & \textbf{Nodes} & \textbf{Edges} & \textbf{Features} & \textbf{Hiddens} & \textbf{Dropout} & \textbf{GCN~(\%)} & \textbf{SAGE~(\%)} & \textbf{GAT~(\%)} \\
        \midrule
        Flickr~\cite{graphsaint_20} & 89~K & 900~K & 500 & 128 & 0.5 & 51.8 & 51.2 & 51.6 \\
        Reddit~\cite{GraphSage2017} & 232~K & 115~M & 602 & 128 & 0.5 & 89.3 & 96.2 & 95.3 \\
        Yelp~\cite{graphsaint_20} & 717~K & 14.0~M & 300 & 512 & 0.1 & 42.8 & 64.0 & 57.1 \\
        Amazon~\cite{graphsaint_20} & 1.6~M & 264~M & 200 & 512 & 0.1 & 41.6 & 79.4 & 63.4 \\
        Products~\cite{OGB2020} & 2.4~M & 124~M & 100 & 128 & 0.5 & 76.2 & 78.3 & 74.1 \\
        Papers~\cite{OGB2020} & 111~M & 1.6~B & 128 & 512 & 0.5 & 47.2 & 50.9 & 49.6 \\
        \bottomrule
    \end{tabular}
}
\caption{
Datasets used in recomputation policy evaluations. The first three columns describe the number of nodes, edges, and feature dimensions, respectively. The next two columns show the hidden dimensions and dropout probabilities used in training GNN models. Finally, the last three columns show the accuracies of the trained models on each test dataset.
}
\label{tab:appendix-dataset}
\end{table}

\begin{figure*}[!t]
\centering
  \begin{subfigure}{.27\textwidth}
    \includegraphics[width=\linewidth]{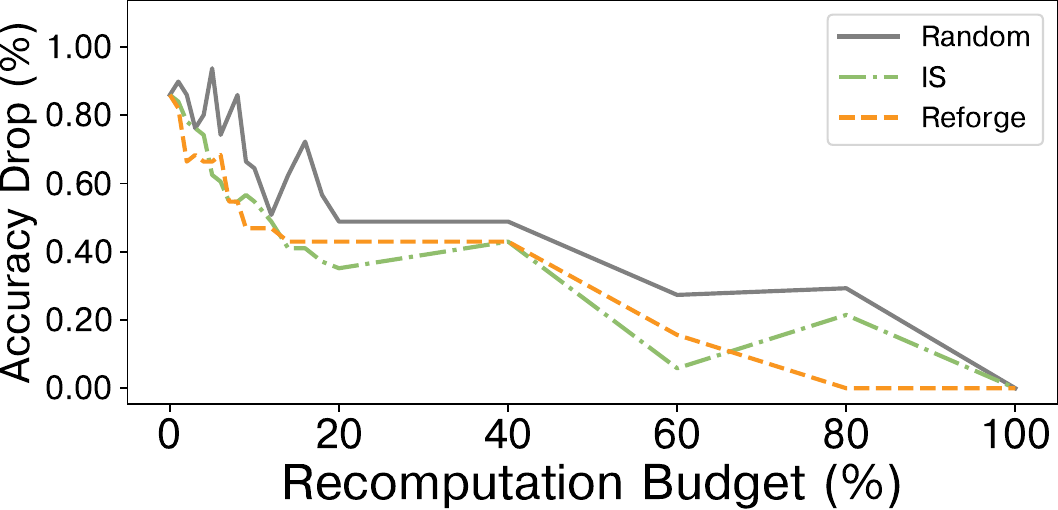}
    \caption{Flickr GCN (52.5\%)}
  \end{subfigure}
  \begin{subfigure}{.27\textwidth}
    \includegraphics[width=\linewidth]{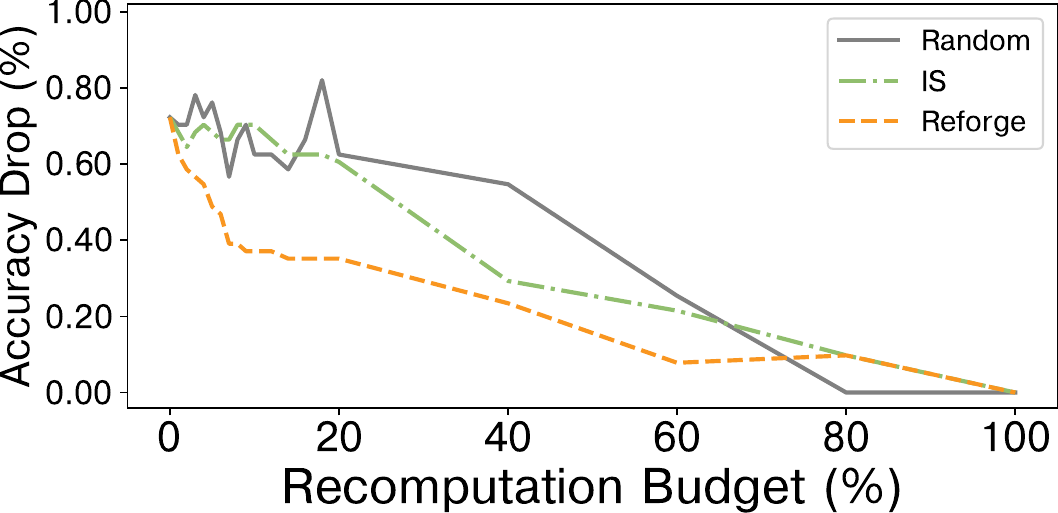}
    \caption{Flickr SAGE (51.2\%)}
  \end{subfigure}
  \begin{subfigure}{.27\textwidth}
    \includegraphics[width=\linewidth]{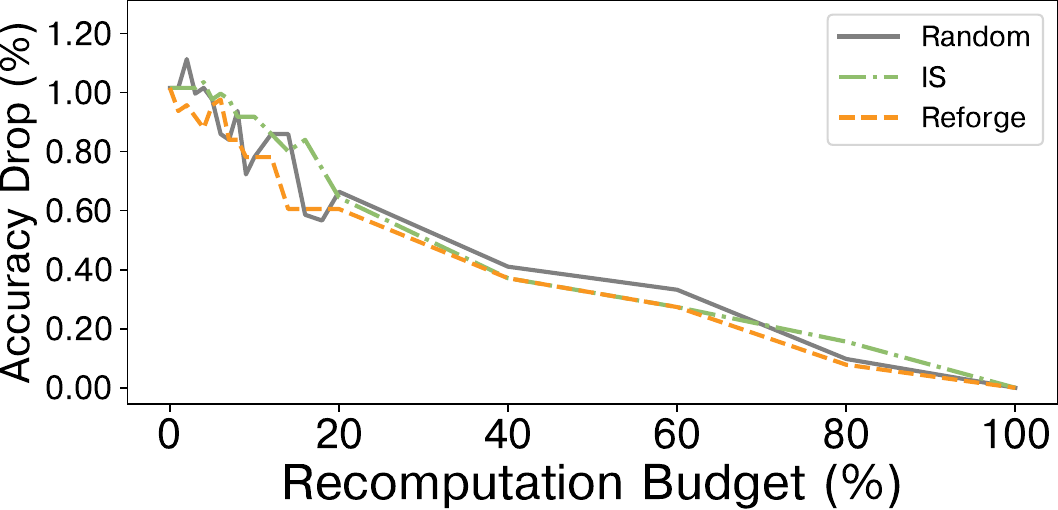}
    \caption{Flickr GAT (51.5\%)}
  \end{subfigure}

  \begin{subfigure}{.27\textwidth}
    \includegraphics[width=\linewidth]{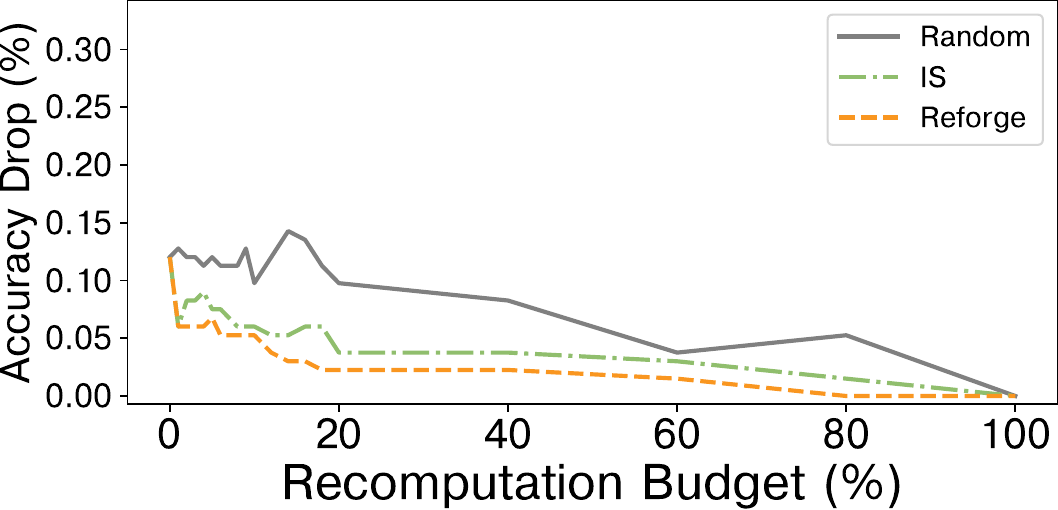}
    \caption{Reddit GCN (89.1\%)}
  \end{subfigure}
  \begin{subfigure}{.27\textwidth}
    \includegraphics[width=\linewidth]{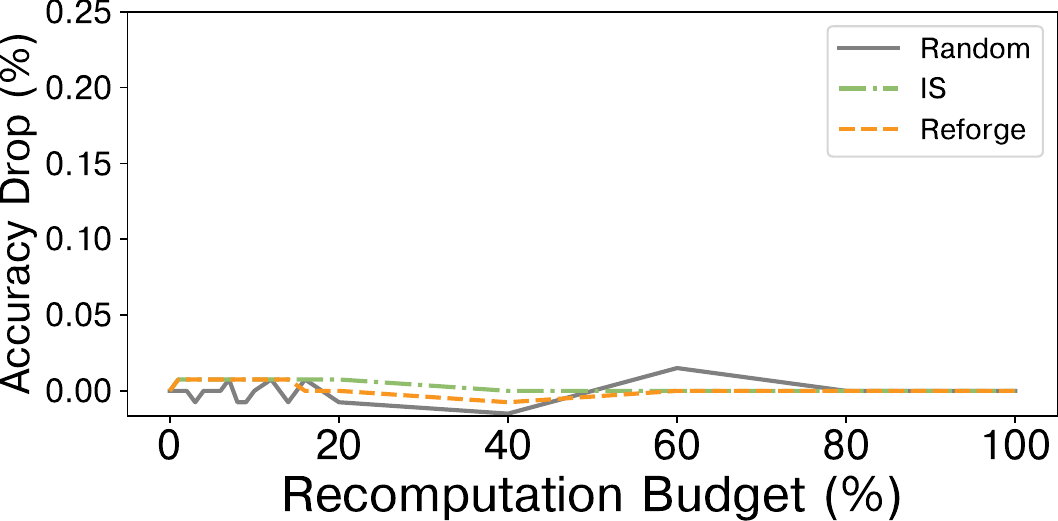}
    \caption{Reddit SAGE (96.0\%)}
  \end{subfigure}
  \begin{subfigure}{.27\textwidth}
    \includegraphics[width=\linewidth]{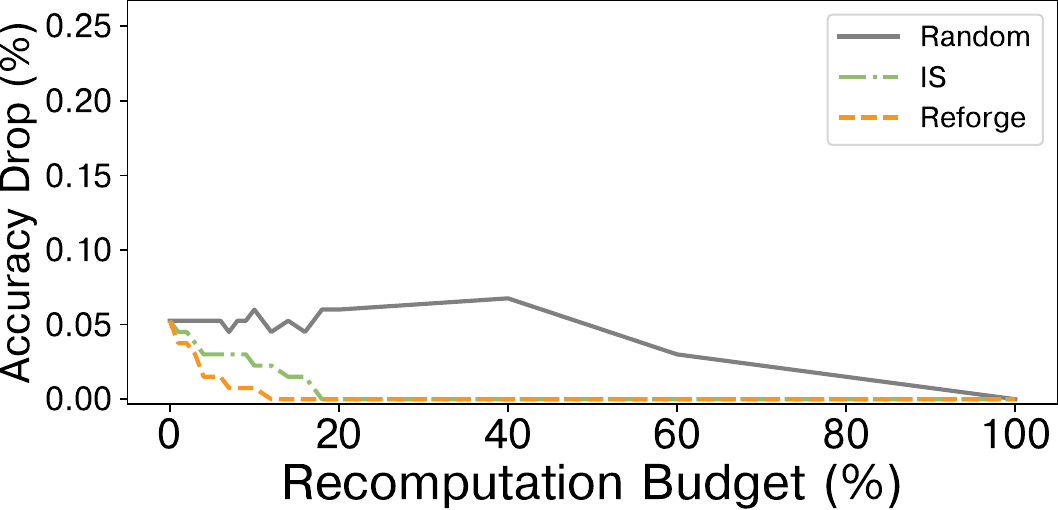}
    \caption{Reddit GAT (95.2\%)}
  \end{subfigure}

  \begin{subfigure}{.27\textwidth}
    \includegraphics[width=\linewidth]{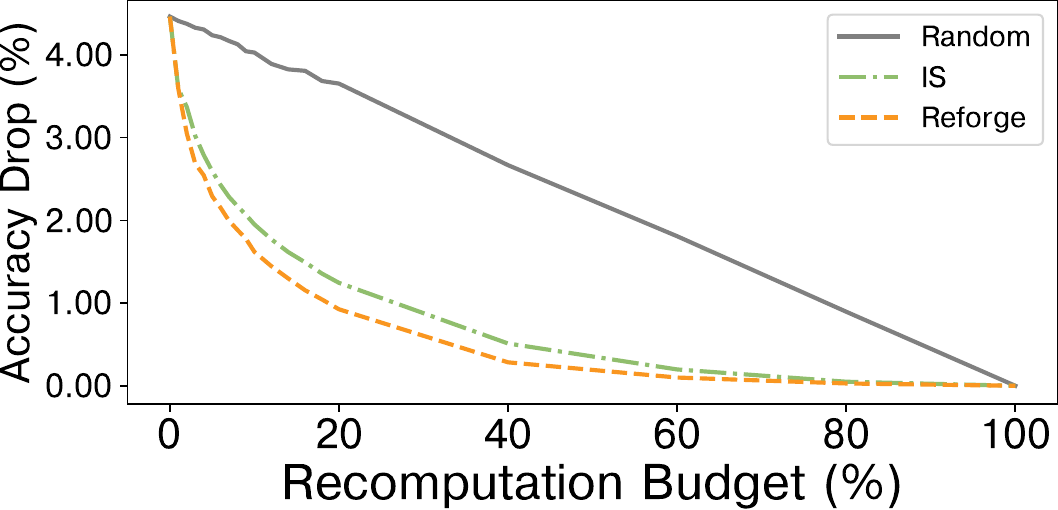}
    \caption{Yelp GCN (42.4\%)}
  \end{subfigure}
  \begin{subfigure}{.27\textwidth}
    \includegraphics[width=\linewidth]{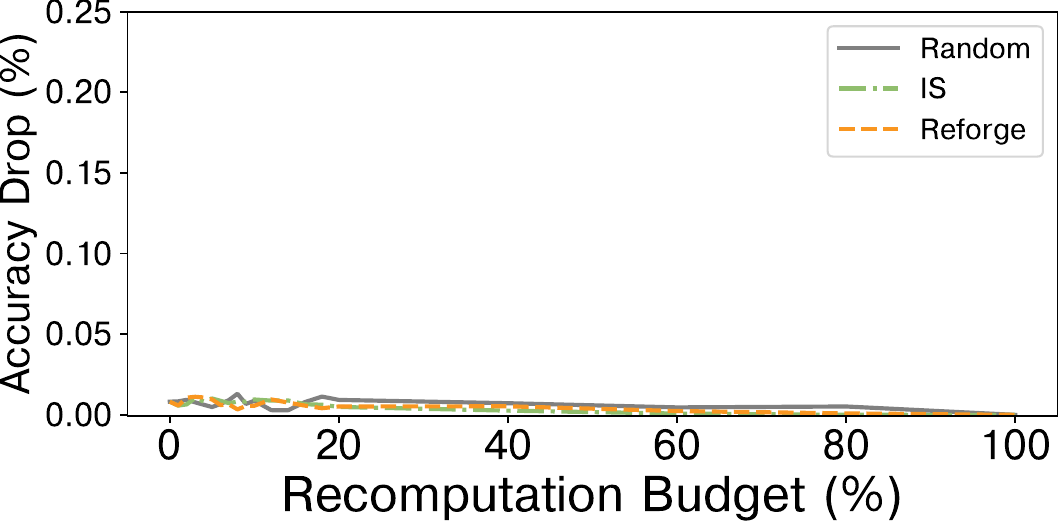}
    \caption{Yelp SAGE (63.7\%)}
  \end{subfigure}
  \begin{subfigure}{.27\textwidth}
    \includegraphics[width=\linewidth]{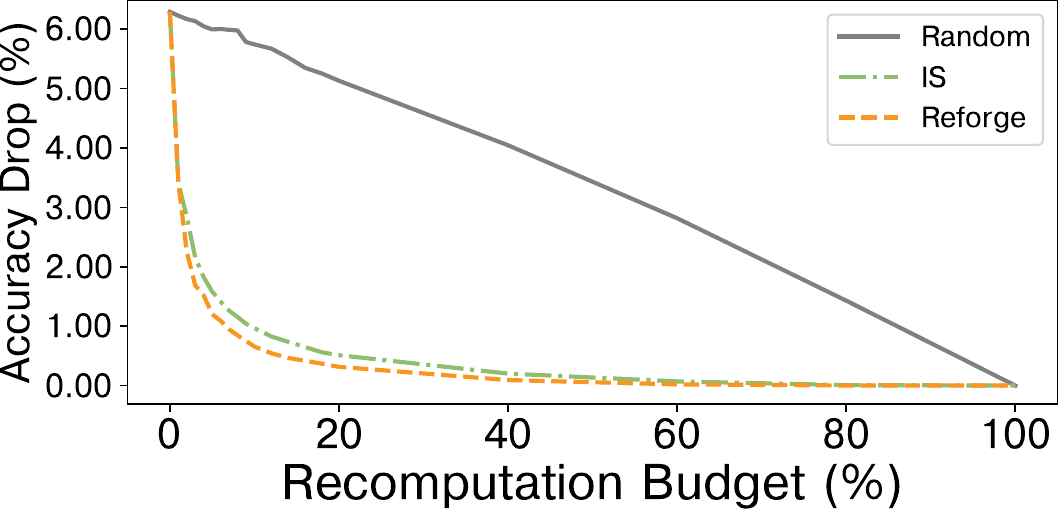}
    \caption{Yelp GAT (56.9\%)}
  \end{subfigure}

  \begin{subfigure}{.27\textwidth}
    \includegraphics[width=\linewidth]{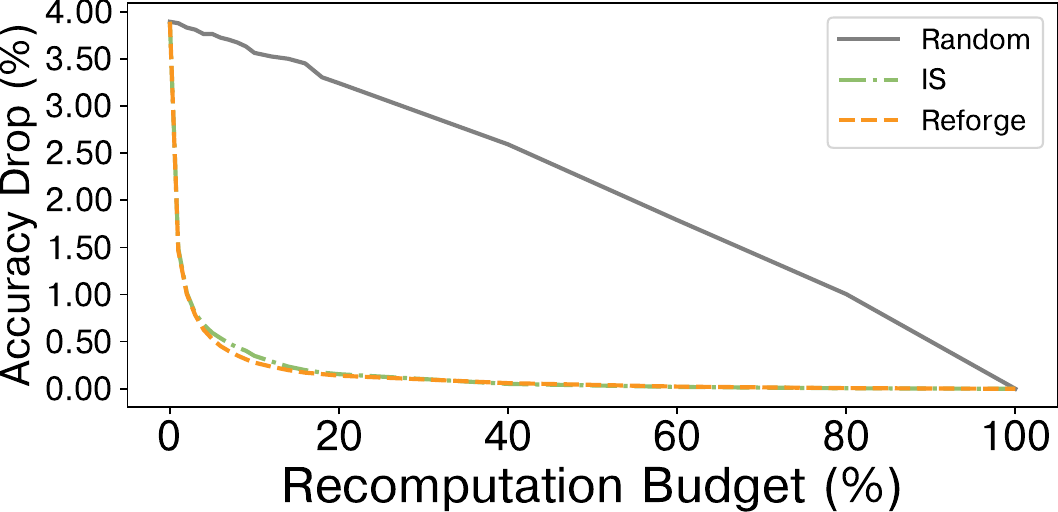}
    \caption{Amazon GCN (41.8\%)}
  \end{subfigure}
  \begin{subfigure}{.27\textwidth}
    \includegraphics[width=\linewidth]{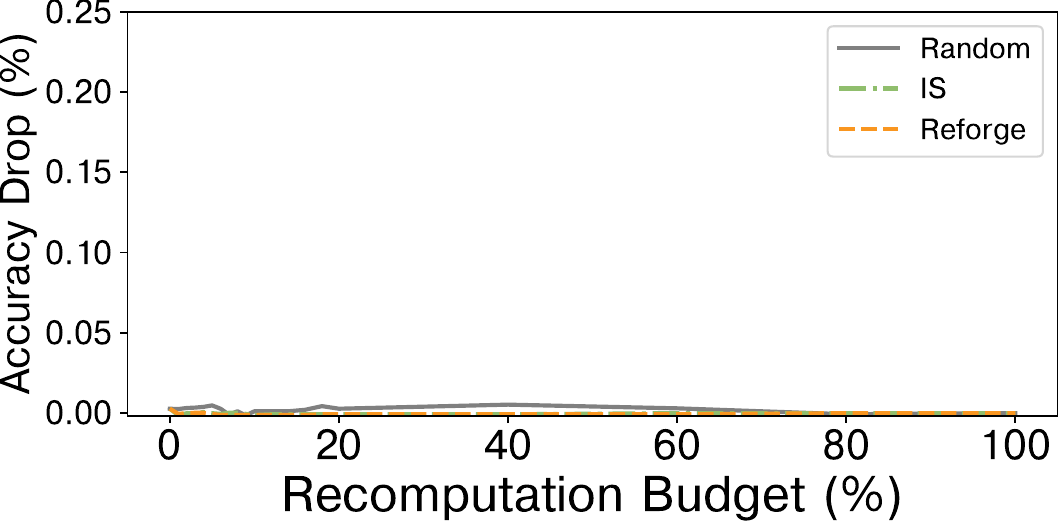}
    \caption{Amazon SAGE (79.4\%)}
  \end{subfigure}
  \begin{subfigure}{.27\textwidth}
    \includegraphics[width=\linewidth]{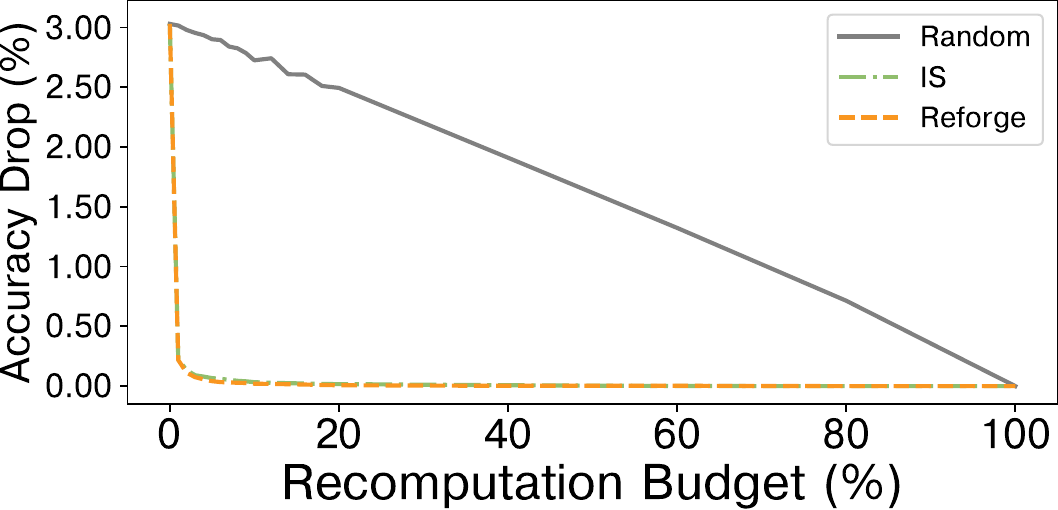}
    \caption{Amazon GAT (63.4\%)}
  \end{subfigure}

  \begin{subfigure}{.27\textwidth}
    \includegraphics[width=\linewidth]{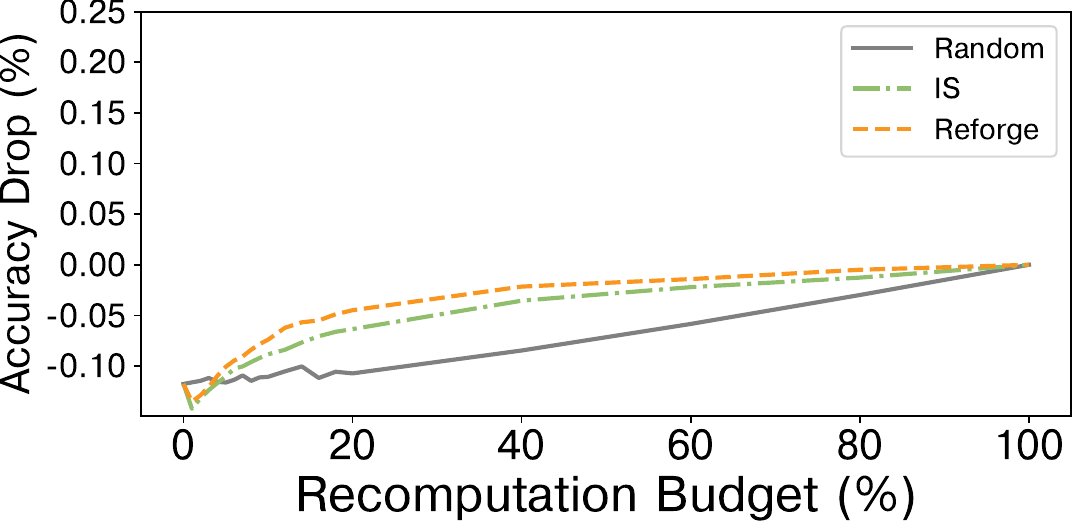}
    \caption{Products GCN (75.6\%)}
  \end{subfigure}
  \begin{subfigure}{.27\textwidth}
    \includegraphics[width=\linewidth]{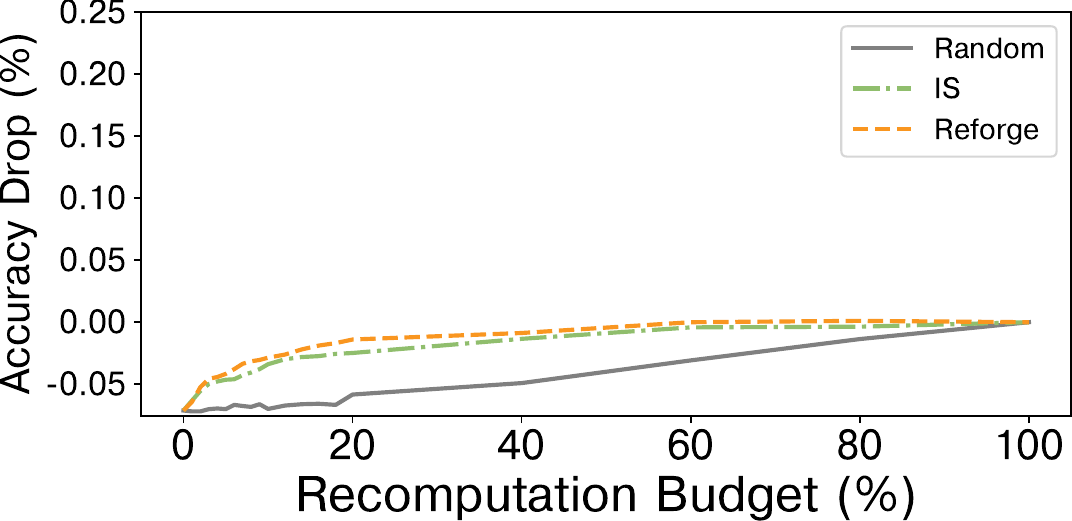}
    \caption{Products SAGE (77.8\%)}
  \end{subfigure}
  \begin{subfigure}{.27\textwidth}
    \includegraphics[width=\linewidth]{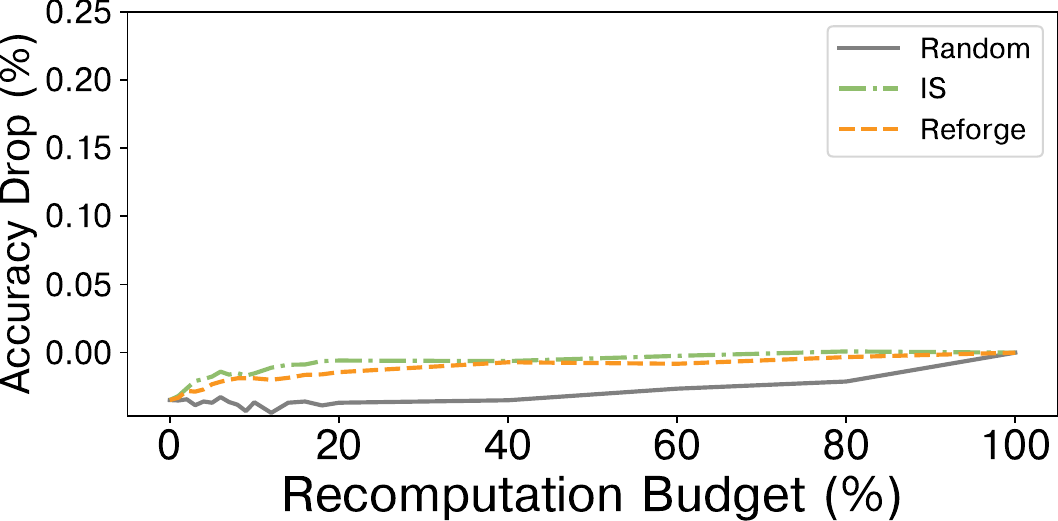}
    \caption{Products GAT (73.3\%)}
  \end{subfigure}

  \begin{subfigure}{.27\textwidth}
    \includegraphics[width=\linewidth]{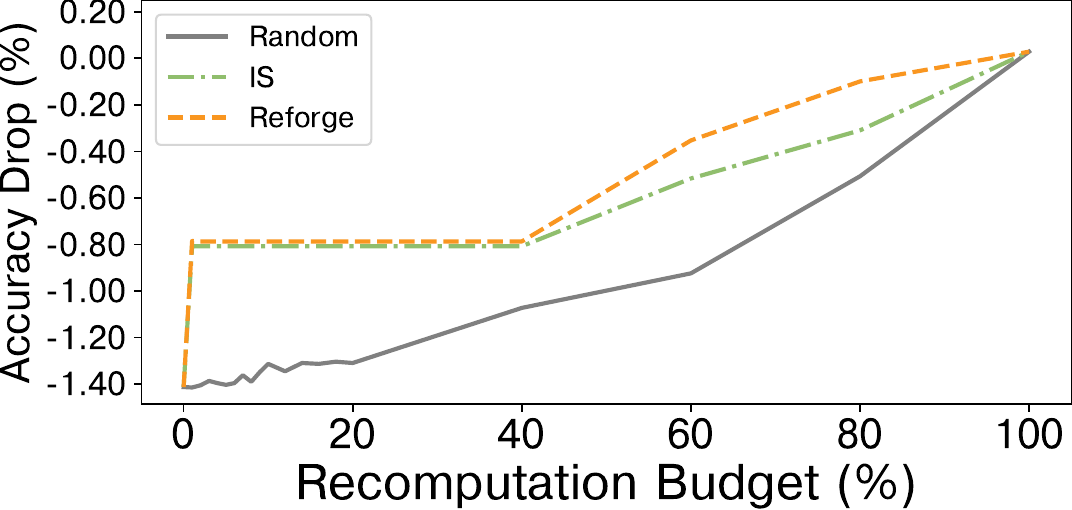}
    \caption{Papers GCN (42.0\%)}
  \end{subfigure}
  \begin{subfigure}{.27\textwidth}
    \includegraphics[width=\linewidth]{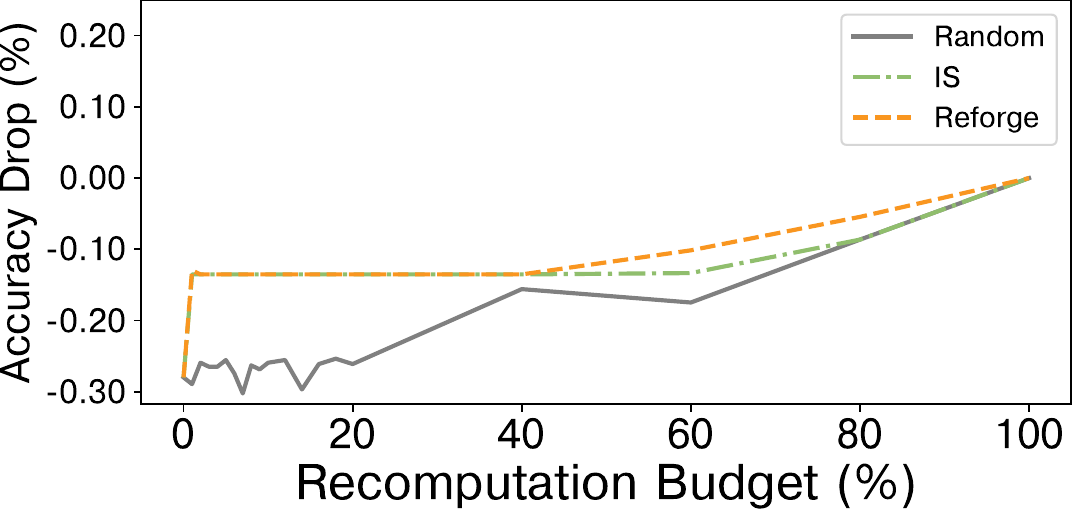}
    \caption{Papers SAGE (50.0\%)}
  \end{subfigure}
  \begin{subfigure}{.27\textwidth}
    \includegraphics[width=\linewidth]{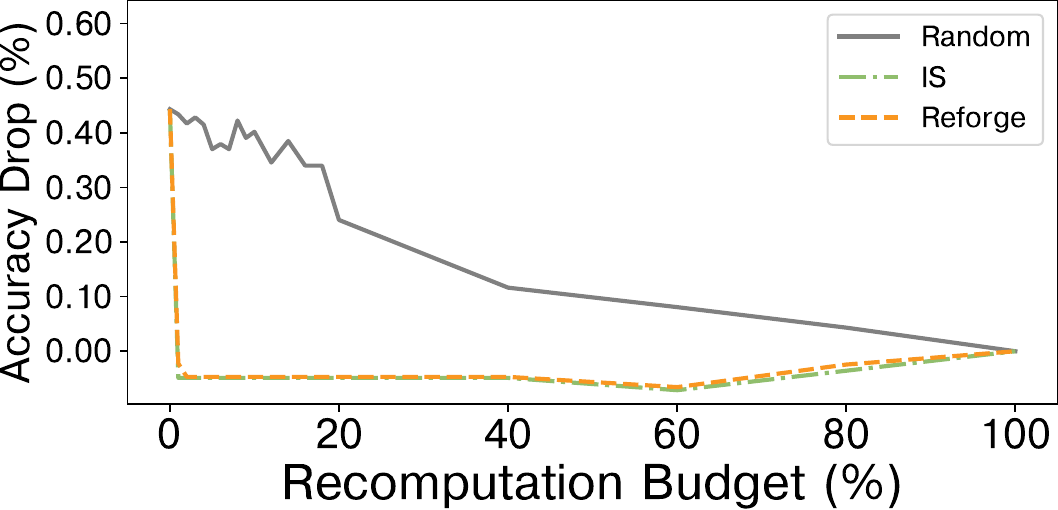}
    \caption{Papers GAT (49.3\%)}
  \end{subfigure}
  \caption{
  The recovered accuracy varies under different recomputation policies. Each figure's caption lists the graph dataset, GNN model, and accuracy with full computation graphs. To evaluate the recomputation policies (\ie \textsf{Random}, \textsf{IS}, and \sys) we randomly remove 25\% of test nodes to create batches, each with 1,024 query nodes, and compute PEs using the remaining nodes. Accuracy drops are then aggregated across these batches. We describe the dataset profiles in \autoref{tab:appendix-dataset}.  Note that the model performance reported in~\autoref{tab:appendix-dataset} is measured with the entire test nodes, which can differ from the accuracy with full computation graphs here.
  }
  \label{fig:appendix_srpe_recoms}
\end{figure*}

\section{Extended Evaluation on Recomputation Policies}
\label{appendix:srpe_eval}

In this section, we extend the evaluation in \autoref{subsubsec:policy} to more thoroughly assess the performance of \sys's recomputation policy compared to the \textsf{Random} and \textsf{IS} policies. We train three representative GNN models—GCN, SAGE, and GAT—on six different datasets, as detailed in \autoref{tab:appendix-dataset}. After training, we remove 25\% of test nodes (and their connected edges) at random, forming batches of size 1,024. We compute PEs for the training nodes and the remaining 75\% of test nodes for each model-dataset pair. We assess the performance of recomputation policies using these batches and PEs and aggregate the results.
\autoref{fig:appendix_srpe_recoms} shows that \sys's recomputation policy outperforms the baseline policies in most cases, effectively narrowing the accuracy gap between full computation graphs and those utilizing SRPE.

\end{document}